\documentclass[aps,prl,twocolumn,showpacs,groupedaddress,floatfix,superscriptaddress]{revtex4} 
\usepackage{graphicx}
\usepackage{dcolumn}
\usepackage{bm}
\usepackage{amssymb}
\usepackage{multirow}
\usepackage{verbatim}
\usepackage{array}

\def\ifb{fb$^{-1}$}
\def\pp{$p\bar{p}$}
\def\tevE{$\sqrt{s}=1.96$~TeV}
\def\wjets{$W+$jets}

\def\ttbar{$t\bar{t}$}
\def\met{\ensuremath{E\kern-0.57em/_{T}}}

\def\wwnlo{11.7}
\def\wwnloe{0.8}
\def\wznlo{3.5}	
\def\wznloe{0.3}
\def\zznlo{1.4} 
\def\zznloe{0.1}

\def\wzRF{3.3}
\def\wzRFeu{4.1}
\def\wzRFed{3.3}

\def\wvRFdchi{63.5}
\def\wvRFsd{7.9}

\def\wvRFdchiExp{35.8}
\def\wvRFsdexp{5.9}

\def\wvMJJdchi{33.0}
\def\wvMJJsd{5.6}

\def\wvMJJdchiExp{22.1}
\def\wvMJJsdexp{4.6}

\begin{document}

  \hspace{5.2in} \mbox{Fermilab-Pub-11/635-E}

  \title{\boldmath Measurements of $WW$ and $WZ$ production in $W$ + jets final states in \\
  $p\bar{p}$ collisions} 

  \affiliation{Universidad de Buenos Aires, Buenos Aires, Argentina}
\affiliation{LAFEX, Centro Brasileiro de Pesquisas F{\'\i}sicas, Rio de Janeiro, Brazil}
\affiliation{Universidade do Estado do Rio de Janeiro, Rio de Janeiro, Brazil}
\affiliation{Universidade Federal do ABC, Santo Andr\'e, Brazil}
\affiliation{Instituto de F\'{\i}sica Te\'orica, Universidade Estadual Paulista, S\~ao Paulo, Brazil}
\affiliation{University of Science and Technology of China, Hefei, People's Republic of China}
\affiliation{Universidad de los Andes, Bogot\'{a}, Colombia}
\affiliation{Charles University, Faculty of Mathematics and Physics, Center for Particle Physics, Prague, Czech Republic}
\affiliation{Czech Technical University in Prague, Prague, Czech Republic}
\affiliation{Center for Particle Physics, Institute of Physics, Academy of Sciences of the Czech Republic, Prague, Czech Republic}
\affiliation{Universidad San Francisco de Quito, Quito, Ecuador}
\affiliation{LPC, Universit\'e Blaise Pascal, CNRS/IN2P3, Clermont, France}
\affiliation{LPSC, Universit\'e Joseph Fourier Grenoble 1, CNRS/IN2P3, Institut National Polytechnique de Grenoble, Grenoble, France}
\affiliation{CPPM, Aix-Marseille Universit\'e, CNRS/IN2P3, Marseille, France}
\affiliation{LAL, Universit\'e Paris-Sud, CNRS/IN2P3, Orsay, France}
\affiliation{LPNHE, Universit\'es Paris VI and VII, CNRS/IN2P3, Paris, France}
\affiliation{CEA, Irfu, SPP, Saclay, France}
\affiliation{IPHC, Universit\'e de Strasbourg, CNRS/IN2P3, Strasbourg, France}
\affiliation{IPNL, Universit\'e Lyon 1, CNRS/IN2P3, Villeurbanne, France and Universit\'e de Lyon, Lyon, France}
\affiliation{III. Physikalisches Institut A, RWTH Aachen University, Aachen, Germany}
\affiliation{Physikalisches Institut, Universit{\"a}t Freiburg, Freiburg, Germany}
\affiliation{II. Physikalisches Institut, Georg-August-Universit{\"a}t G\"ottingen, G\"ottingen, Germany}
\affiliation{Institut f{\"u}r Physik, Universit{\"a}t Mainz, Mainz, Germany}
\affiliation{Ludwig-Maximilians-Universit{\"a}t M{\"u}nchen, M{\"u}nchen, Germany}
\affiliation{Fachbereich Physik, Bergische Universit{\"a}t Wuppertal, Wuppertal, Germany}
\affiliation{Panjab University, Chandigarh, India}
\affiliation{Delhi University, Delhi, India}
\affiliation{Tata Institute of Fundamental Research, Mumbai, India}
\affiliation{University College Dublin, Dublin, Ireland}
\affiliation{Korea Detector Laboratory, Korea University, Seoul, Korea}
\affiliation{CINVESTAV, Mexico City, Mexico}
\affiliation{Nikhef, Science Park, Amsterdam, the Netherlands}
\affiliation{Radboud University Nijmegen, Nijmegen, the Netherlands and Nikhef, Science Park, Amsterdam, the Netherlands}
\affiliation{Joint Institute for Nuclear Research, Dubna, Russia}
\affiliation{Institute for Theoretical and Experimental Physics, Moscow, Russia}
\affiliation{Moscow State University, Moscow, Russia}
\affiliation{Institute for High Energy Physics, Protvino, Russia}
\affiliation{Petersburg Nuclear Physics Institute, St. Petersburg, Russia}
\affiliation{Instituci\'{o} Catalana de Recerca i Estudis Avan\c{c}ats (ICREA) and Institut de F\'{i}sica d'Altes Energies (IFAE), Barcelona, Spain}
\affiliation{Stockholm University, Stockholm and Uppsala University, Uppsala, Sweden}
\affiliation{Lancaster University, Lancaster LA1 4YB, United Kingdom}
\affiliation{Imperial College London, London SW7 2AZ, United Kingdom}
\affiliation{The University of Manchester, Manchester M13 9PL, United Kingdom}
\affiliation{University of Arizona, Tucson, Arizona 85721, USA}
\affiliation{University of California Riverside, Riverside, California 92521, USA}
\affiliation{Florida State University, Tallahassee, Florida 32306, USA}
\affiliation{Fermi National Accelerator Laboratory, Batavia, Illinois 60510, USA}
\affiliation{University of Illinois at Chicago, Chicago, Illinois 60607, USA}
\affiliation{Northern Illinois University, DeKalb, Illinois 60115, USA}
\affiliation{Northwestern University, Evanston, Illinois 60208, USA}
\affiliation{Indiana University, Bloomington, Indiana 47405, USA}
\affiliation{Purdue University Calumet, Hammond, Indiana 46323, USA}
\affiliation{University of Notre Dame, Notre Dame, Indiana 46556, USA}
\affiliation{Iowa State University, Ames, Iowa 50011, USA}
\affiliation{University of Kansas, Lawrence, Kansas 66045, USA}
\affiliation{Kansas State University, Manhattan, Kansas 66506, USA}
\affiliation{Louisiana Tech University, Ruston, Louisiana 71272, USA}
\affiliation{Boston University, Boston, Massachusetts 02215, USA}
\affiliation{Northeastern University, Boston, Massachusetts 02115, USA}
\affiliation{University of Michigan, Ann Arbor, Michigan 48109, USA}
\affiliation{Michigan State University, East Lansing, Michigan 48824, USA}
\affiliation{University of Mississippi, University, Mississippi 38677, USA}
\affiliation{University of Nebraska, Lincoln, Nebraska 68588, USA}
\affiliation{Rutgers University, Piscataway, New Jersey 08855, USA}
\affiliation{Princeton University, Princeton, New Jersey 08544, USA}
\affiliation{State University of New York, Buffalo, New York 14260, USA}
\affiliation{Columbia University, New York, New York 10027, USA}
\affiliation{University of Rochester, Rochester, New York 14627, USA}
\affiliation{State University of New York, Stony Brook, New York 11794, USA}
\affiliation{Brookhaven National Laboratory, Upton, New York 11973, USA}
\affiliation{Langston University, Langston, Oklahoma 73050, USA}
\affiliation{University of Oklahoma, Norman, Oklahoma 73019, USA}
\affiliation{Oklahoma State University, Stillwater, Oklahoma 74078, USA}
\affiliation{Brown University, Providence, Rhode Island 02912, USA}
\affiliation{University of Texas, Arlington, Texas 76019, USA}
\affiliation{Southern Methodist University, Dallas, Texas 75275, USA}
\affiliation{Rice University, Houston, Texas 77005, USA}
\affiliation{University of Virginia, Charlottesville, Virginia 22901, USA}
\affiliation{University of Washington, Seattle, Washington 98195, USA}
\author{V.M.~Abazov} \affiliation{Joint Institute for Nuclear Research, Dubna, Russia}
\author{B.~Abbott} \affiliation{University of Oklahoma, Norman, Oklahoma 73019, USA}
\author{B.S.~Acharya} \affiliation{Tata Institute of Fundamental Research, Mumbai, India}
\author{M.~Adams} \affiliation{University of Illinois at Chicago, Chicago, Illinois 60607, USA}
\author{T.~Adams} \affiliation{Florida State University, Tallahassee, Florida 32306, USA}
\author{G.D.~Alexeev} \affiliation{Joint Institute for Nuclear Research, Dubna, Russia}
\author{G.~Alkhazov} \affiliation{Petersburg Nuclear Physics Institute, St. Petersburg, Russia}
\author{A.~Alton$^{a}$} \affiliation{University of Michigan, Ann Arbor, Michigan 48109, USA}
\author{G.~Alverson} \affiliation{Northeastern University, Boston, Massachusetts 02115, USA}
\author{G.A.~Alves} \affiliation{LAFEX, Centro Brasileiro de Pesquisas F{\'\i}sicas, Rio de Janeiro, Brazil}
% ...
%
%
\collaboration{The D0 Collaboration\footnote{with visitors from
$^{a}$Augustana College, Sioux Falls, SD, USA,
$^{b}$The University of Liverpool, Liverpool, UK,
$^{c}$UPIITA-IPN, Mexico City, Mexico,
$^{d}$SLAC, Menlo Park, CA, USA,
$^{e}$University College London, London, UK,
$^{f}$Centro de Investigacion en Computacion - IPN, Mexico City, Mexico,
$^{g}$ECFM, Universidad Autonoma de Sinaloa, Culiac\'an, Mexico,
and 
$^{h}$Universit{\"a}t Bern, Bern, Switzerland.
$^{\ddag}$Deceased.
}} \noaffiliation
\vskip 0.25cm

  \date{December 1, 2011}
  
  \begin{abstract}
    We study $WW$ and $WZ$ production with $\ell\nu q{q}$ ($\ell=e,\mu$) 
    final states using data collected by the D0 detector at the Fermilab 
    Tevatron Collider corresponding to 4.3~\ifb\ of integrated luminosity 
    from \pp\ collisions at \tevE.  Assuming the ratio between the production 
    cross sections $\sigma(WW)$ and $\sigma(WZ)$ as predicted by the 
    standard model, we measure the total $WV$ ($V=W,Z$) cross section to be 
    $\sigma(WV)= 19.6~^{+3.2}_{-3.0}$~pb, and reject the background-only 
    hypothesis at a level of 7.9 standard deviations.  We also use $b$-jet 
    discrimination to separate the $WZ$ component from the dominant $WW$ 
    component.  Simultaneously fitting $WW$ and $WZ$ contributions, we measure 
    $\sigma(WW)=15.9~^{+3.7}_{-3.2}$~pb and $\sigma(WZ)=\wzRF~^{+\wzRFeu}_{-\wzRFed}$~pb, 
    which is consistent with the standard model predictions.
  \end{abstract}
  
  \maketitle

  The study of the production of $VV$ ($V=W,Z$) boson pairs provides 
  an important test of the electroweak sector of the standard model 
  (SM).  In $p\bar{p}$ collisions at $\sqrt{s}=1.96$ TeV, the 
  next-to-leading order (NLO) SM cross sections for these processes are
  $\sigma(WW)=\wwnlo\pm\wwnloe$~pb, $\sigma(WZ)=\wznlo\pm\wznloe$~pb
  and $\sigma(ZZ)=\zznlo\pm\zznloe$~pb~\cite{bib:Campbell}.  Measuring
  a significant departure in cross section or deviations in the
  predicted kinematic distributions would indicate the presence of
  anomalous gauge boson couplings~\cite{bib:anocoups} or new particles
  in extensions of the SM~\cite{bib:newphen}.  This analysis also
  provides a proving ground for the advanced analysis techniques used
  in low mass Higgs searches~\cite{bib:higgs}.  The production of $VV$
  in $p\bar{p}$ collisions at the Fermilab Tevatron Collider has been
  observed in fully leptonic decay modes~\cite{bib:leptonic} and more
  recently, in leptons+jets decay modes~\cite{bib:hadronic}, where the
  combined $WW+WZ$ cross section was measured.  \\
  \indent In this Letter, we report observation of the production of a $W$
  boson that decays leptonically in associated production with a
  second vector boson that decays hadronically ($WV\rightarrow\ell\nu
  q{q};~\ell=e^{\pm}$ or $\mu^{\pm}$, and $\nu$ and $q$ denote matter
  or anti-matter as appropriate). The data used for this analysis 
  correspond to 4.3~\ifb\ of integrated luminosity collected between 
  2006 and 2009 by the D0 detector~\cite{bib:detector} at the Fermilab 
  Tevatron Collider.  The D0 detector dijet mass resolution for $W/Z$ 
  decays of $\approx$~18\% results in significant overlap of $W\rightarrow q{q}$ 
  and $Z\rightarrow q{q}$ dijet mass peaks.  Therefore, we first consider
  $WW$ and $WZ$ simultaneously and measure the total $WV$ cross
  section assuming the ratio of $WW$ to $WZ$ cross sections as predicted 
  by the SM.  We then apply $b$-jet identification to separate the $WZ$
  contribution, where the $Z$ boson decays into $b\bar{b}$ pairs, from
  the dominant $WW$ production.

  Candidate events in the electron channel are required to satisfy a 
  single electron trigger or a trigger requiring electrons and jets, 
  which results in a combined trigger efficiency of $(98^{+2}_{-3})\%$ 
  for the $e\nu q{q}$ event selection described below.  A comprehensive 
  suite of triggers in the muon channel, based on leptons, jets and their 
  combination, achieves a trigger efficiency of $(95\pm 5)\%$ for the 
  $\mu\nu q{q}$ event selection.

  To select $WV$$\rightarrow\ell\nu q{q}$ candidates, we require a
  single reconstructed electron (muon) with transverse momentum
  $p_T>15$~GeV~(20~GeV) and pseudorapidity $|\eta|<1.1\ (2.0)$~\cite{bib:def}, 
  missing transverse energy $\met>20$~GeV, and two or three jets 
  reconstructed using a cone algorithm~\cite{bib:JetCone}.  The 
  jets must have $p_T>20$~GeV, $|\eta|<2.5$, and at least two tracks 
  within the jet cone~\cite{bib:JetCone} originating from the $p\bar{p}$ 
  interaction vertex.  Lepton candidates must 
  be spatially matched to a track that originates from the primary $p\bar{p}$ 
  interaction vertex and they must be isolated from energy 
  depositions in the calorimeter and other tracks in the central 
  tracking detector.  To reduce background from processes that do 
  not contain $W$$\rightarrow\ell\nu$, we require that the transverse 
  mass~\cite{bib:smithUA1} is $M_T^{\ell\nu}~(\rm{GeV})>40-0.5$\met.  
  In addition, we restrict $M_T^{\mu\nu} < 200$~GeV to suppress muon 
  candidates with poorly measured momenta.

  Signal and most of the background processes are
  modeled with Monte Carlo (MC) simulation.  The signal events are
  generated with {\sc pythia}~\cite{bib:PYTHIA} using \textsc{CTEQ6L1}
  parton distribution functions (PDFs)~\cite{bib:CTEQ6} and include 
  all SM decays.  The fixed-order matrix element (FOME) generator {\sc
  alpgen}~\cite{bib:ALPGEN} with \textsc{CTEQ6L1} PDF is used to
  generate $W$+jets, $Z$+jets, and $t\bar{t}$ events.  The FOME
  generator {\sc comphep}~\cite{bib:CompHEP} is used to produce single
  top-quark MC samples with \textsc{CTEQ6M} PDF~\cite{bib:CTEQ6}.  
  Both {\sc alpgen} and {\sc comphep} are interfaced to {\sc pythia} 
  for parton showering and hadronization.  The MC events undergo a {\sc
  geant}-based~\cite{bib:GEANT} detector simulation and are
  reconstructed using the same algorithms as used for D0 data.  The
  effect of multiple $p\bar{p}$ interactions is included by overlaying
  data events from random beam crossings on simulated events.
  The next-to-NLO (NNLO) cross section is used to normalize the
  $Z$+jets (light and heavy-flavor jets)~\cite{bib:fewz}.  The approximate 
  NNLO cross section~\cite{bib:xsecsTT} is used to normalize the 
  $t\bar{t}$ samples, while the single top-quark MC samples are 
  normalized to the approximate next-to-NNLO cross section~\cite{bib:xsecsT}.  
  The normalization of the $W$+jets MC sample (for all flavor contributions) 
  is determined from data.  Additional NLO heavy-flavor corrections are 
  calculated with {\sc mcfm}~\cite{bib:Campbell2} and applied to 
  $Z/W$+heavy-flavor jets MC samples.

  The multijet background in which a jet is misidentified as a
  prompt lepton is determined from data.  For the muon channel, 
  the multijet background is modeled with data that fail the muon
  isolation requirements, but pass all other selections.  For the 
  electron channel, the multijet background is estimated using a 
  data sample containing events that pass less restrictive electron quality 
  requirements.  Both multijet samples are corrected for contributions 
  from processes modeled by MC.  The multijet normalizations are 
  determined from fits to the $M_T^{\ell\nu}$ distributions
  and assigned uncertainties of 20\%.

  To identify $b$-quark jets, in particular those originating from 
  $Z\rightarrow b\bar{b}$ decays, we use the D0 neural network (NN) 
  $b$-tagging algorithm~\cite{bib:btagging}.  The NN is trained to 
  separate light-flavor jets from heavy-flavor jets based on a 
  combination of variables sensitive to the presence of tracks
  and vertices displaced from the primary $p\bar{p}$ interaction vertex.  
  The NN outputs for the two highest $p_{T}$ jets are then used as inputs 
  to the final multivariate discriminant.  We define non-overlapping 0, 
  1, and 2-tag sub-channels based on whether neither, only one, or both 
  of the two highest $p_T$ jets pass the least restrictive NN operating point, 
  for which the $b$-jet identification efficiency and the light-flavor 
  jet misidentification rate are approximately 80\% and 10\%.  Scale 
  factors are applied to the MC events to account for any difference in 
  efficiency or misidentification rate between data and simulation.

  The dominant background is \wjets\ and therefore the modeling of this 
  process in {\sc alpgen} and the corresponding sources of uncertainties
  were studied in detail.  Comparison of {\sc alpgen} with other 
  generators~\cite{bib:ALPGENcomp} and with data shows discrepancies in 
  jet $\eta$, dijet angular separation and the transverse momentum of the 
  $W$ boson candidate.  Thus, data are used to correct these quantities 
  in the {\sc alpgen} $W+$jets and $Z+$jets samples before $b$-tagging is 
  performed~\cite{bib:d0bump}.  The possible bias in this procedure from 
  the presence of the diboson signal in data is small, but is taken into 
  account as a systematic uncertainty.

  As the diboson events are generated with a LO generator, changes to
  the event kinematics and the acceptance due to a NLO and resummation
  effects are studied using events from the {\sc mc@nlo}~\cite{bib:mc@nlo} 
  interfaced to {\sc herwig}~\cite{bib:herwig} for parton showering and 
  hadronization and using the \textsc{CTEQ6M} PDF set.  Comparing kinematics 
  at the generator level after final state radiation, we parameterize a
  two-dimensional correction matrix in the $p_T$ of the diboson
  system and of the highest $p_{T}$ boson.  After applying this
  correction to our {\sc pythia} sample, we find good agreement with
  {\sc mc@nlo} for all distributions studied.  Half of the difference between 
  the {\sc pythia} and {\sc mc\@nlo} predictions is used as systematic 
  uncertainty on the diboson production model, accounting for the possible 
  effects of higher order corrections beyond NLO and of different showering 
  scenarios.

  The signal and the backgrounds are further separated using a
  multivariate classifier to combine information from several
  variables.  This analysis uses a random forest (RF)
  classifier~\cite{bib:SPR1,bib:SPR2}, from which the output
  distribution is used as a final variable to measure the 
  production cross sections by performing a template fit.  Fifteen 
  well-modeled variables~\cite{bib:EPAPS} that demonstrate a difference 
  in probability density between signal and at least one of the
  backgrounds are used as inputs to the RF.  The RF is trained using 
  a fraction of each MC sample.  The remainder of each MC sample, 
  along with the multijet background samples, is then evaluated by 
  the RF and used in the measurement.
  
    \begin{table}[tbp] 

    \caption{Number of events for signal and each background after the
    combined fit of $WV$ using the RF output distribution (with total
    uncertainties determined from the fit) and the number of events 
    observed in data.}
    
    \label{tab:yields}
    \begin{ruledtabular}
      \begin{tabular}{l 
      @{\extracolsep{\fill}} r 
      @{\extracolsep{3mm}} r @{$\ \pm\ $\extracolsep{0cm}} l 
      @{\extracolsep{\fill}} r 
      @{\extracolsep{2mm}} r @{$\ \pm\ $\extracolsep{0cm}} l}
                   & \multicolumn{3}{c}{Electron channel} & \multicolumn{3}{c}{Muon channel} \\
	\hline	   
	Diboson signal          && 1725  & 84   && 1465  & 67 \\
	$W/Z$+light-flavor jets && 37232 & 1033 && 33516 & 709 \\
	$W/Z$+heavy-flavor jets && 5371  & 608  && 4854  & 490 \\
	\ttbar\ and single top  && 1746  & 127  && 1214  & 86 \\
	Multijet                && 10630 & 1007 && 1982  & 384  \\
        \hline  
	Total predicted         && 56704 & 635  && 43031 & 531 \\ 
	Data                  & \multicolumn{3}{c}{56698} & \multicolumn{3}{c}{43044} \\
      \end{tabular}
    \end{ruledtabular}
  \end{table}

  \begin{figure}[tbp] 
    \begin{centering}
      \includegraphics[width=3.4in]{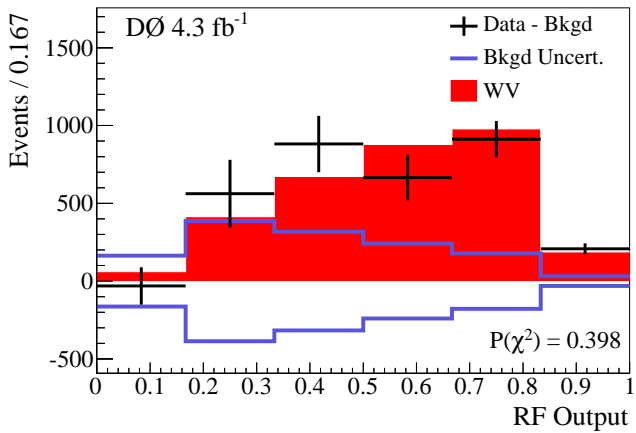}

      \caption{(color online) A comparison of the measured $WV$ signal
	(filled histogram) to background-subtracted data (points) in
	the RF output distribution (summed over electron and muon
	channels, and 0, 1, and 2-tag sub-channels), after the
	combined fit to data using the RF output distributions.  Also
	shown is the $\pm$1 standard deviation uncertainty on the
	background prediction. The $\chi^{2}$ fit probability, 
        P$(\chi^{2})$, is based on the residuals using data and MC 
        statistical uncertainties.}
      
      \label{fig:Fig1}
    \end{centering}
  \end{figure}

  Depending on the source, we consider the effect of systematic 
  uncertainty on the normalization and/or on the shape of differential 
  distributions for signal and backgrounds~\cite{bib:EPAPS}.
  Systematic effects on the differential distributions of the {\sc alpgen} 
  $W$+jets and $Z$+jets MC events from changes of the renormalization 
  and factorization scales and of the parameters used in the MLM 
  parton-jet matching algorithm~\cite{bib:MLM} are also considered.  
  Uncertainties on PDFs~\cite{bib:cteqer}, as well as uncertainties 
  from object reconstruction and identification, are evaluated for 
  all MC samples.

  The total $WV$ cross section is determined from a fit to the data of
  the signal and background RF output distributions.  This fit is
  performed simultaneously on the distributions in the electron and
  muon channels, and in the 0, 1, and 2-tag sub-channels, by
  minimizing a Poisson $\chi^2$ function with respect to variations in
  the systematic uncertainties~\cite{bib:poisson}.  The magnitude of
  the systematic uncertainties is effectively constrained by the
  regions of the RF output distribution with low signal over
  background ratio.  A Gaussian prior is used for each systematic
  uncertainty.  The effects on separate samples or sub-channels due to
  the same uncertainty are assumed to be 100\% correlated.  However,
  different uncertainties are assumed to be mutually independent.

  The fit simultaneously varies the signal and \wjets\ contributions,
  thereby also determining the normalization factor for the \wjets\ MC
  sample.  This obviates the need for using the predicted {\sc alpgen}
  cross section, and provides a more rigorous approach that incorporates 
  an unbiased uncertainty from \wjets\ when extracting the signal cross 
  section.  The \wjets\ normalization factor from the fit is consistent 
  with the theoretical NNLO prediction~\cite{bib:ZWprod}.  
  The yields for signal and each background are given in Table~\ref{tab:yields}.  
  Though the total diboson yield includes a small contribution from 
  $ZZ\to\ell\ell q{q}$ events (1.5\%), in which one of the charged leptons 
  escapes detection, the cross sections presented here are corrected 
  for this contribution assuming that the ratios between $WW$, $WZ$ and $ZZ$ 
  cross sections are given by the SM.

  \begin{figure}[tbp] 
    \begin{centering}
      \includegraphics[width=3.4in]{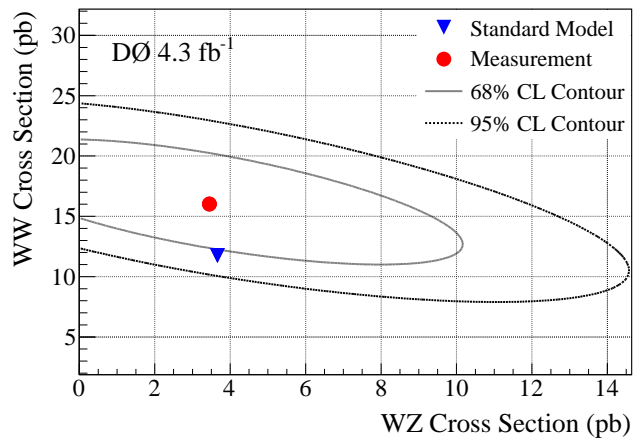}

      \caption{(color online) Results from the simultaneous fit of $\sigma(WW)$ and
      $\sigma(WZ)$ using the RF output distributions.  The plot shows
      the best fit value with 68\% and 95\% confidence level (CL) regions and
      the NLO SM prediction. }
      
      \label{fig:Fig2}
    \end{centering}
  \end{figure}

  \begin{figure*}[htbp] 
    \begin{centering}
      \includegraphics[width=2.3in]{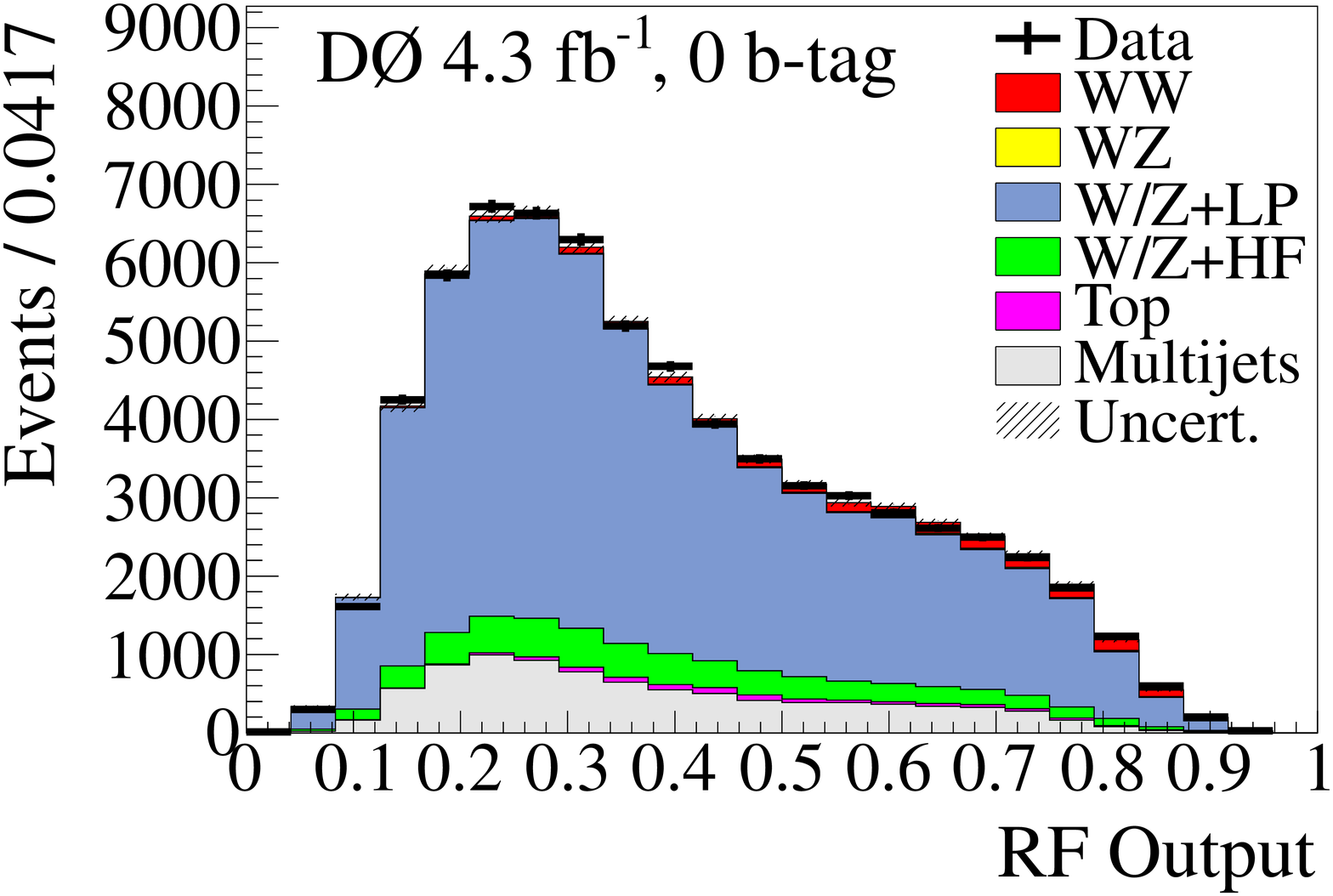}
      \includegraphics[width=2.3in]{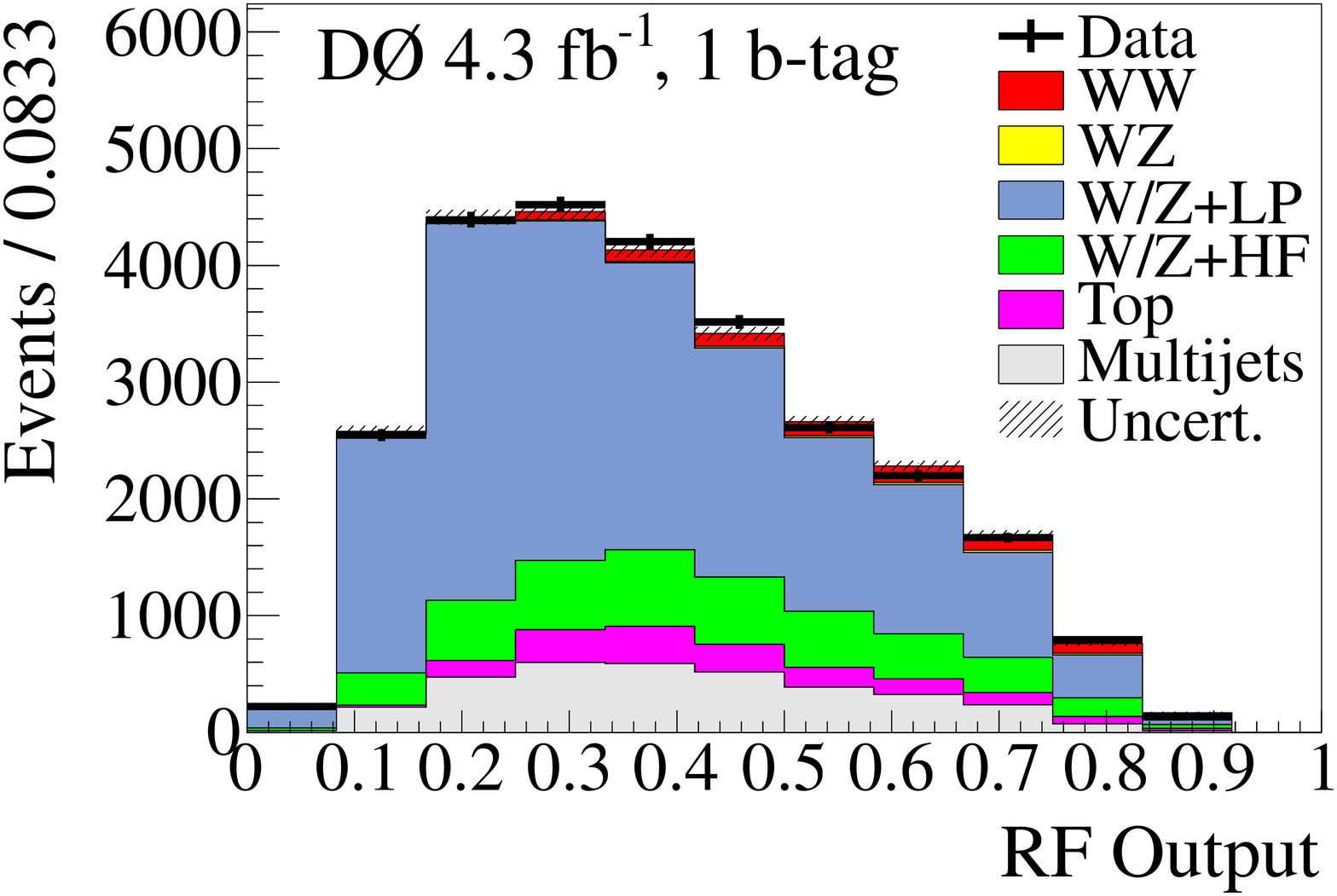}
      \includegraphics[width=2.3in]{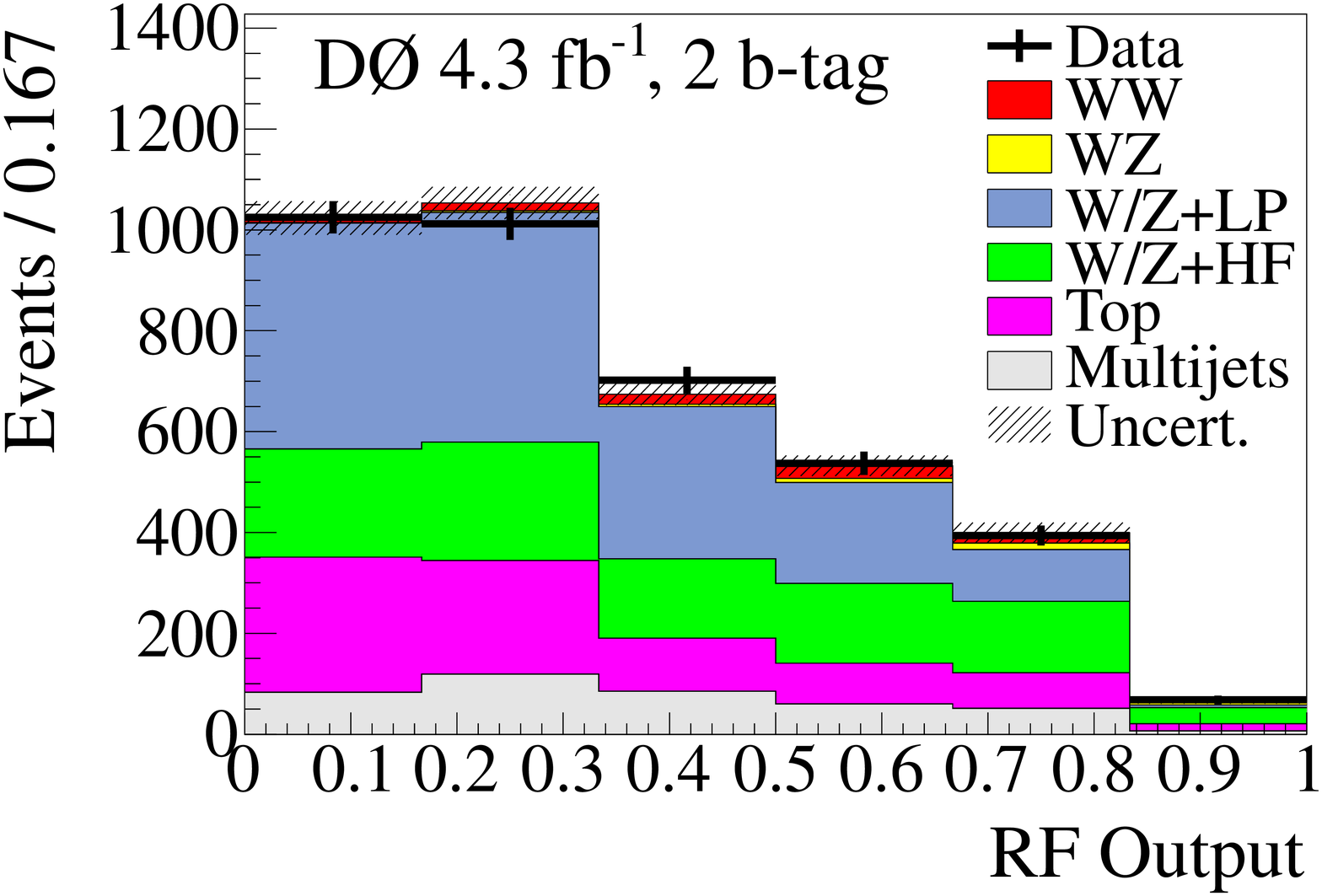}

      \caption{(color online) A comparison of the signal+background prediction to
	data in the RF output distribution (summed over electron and muon channels) 
        for 0, 1, and 2-tag sub-channels after the combined fit to data using 
        the RF output distribution (LP denotes light partons such as $u$, $d$, $s$ 
        or gluon, and HF denotes heavy-flavor such as $c\bar{c}$ or $b\bar{b}$). 
        The systematic uncertainty band is evaluated after the fit of the total 
        $WV$ cross section in the RF output distribution.}
      
      \label{fig:Fig3}
    \end{centering}
  \end{figure*}

  \begin{figure*}[tbp] 
    \begin{centering}
      \includegraphics[width=2.3in]{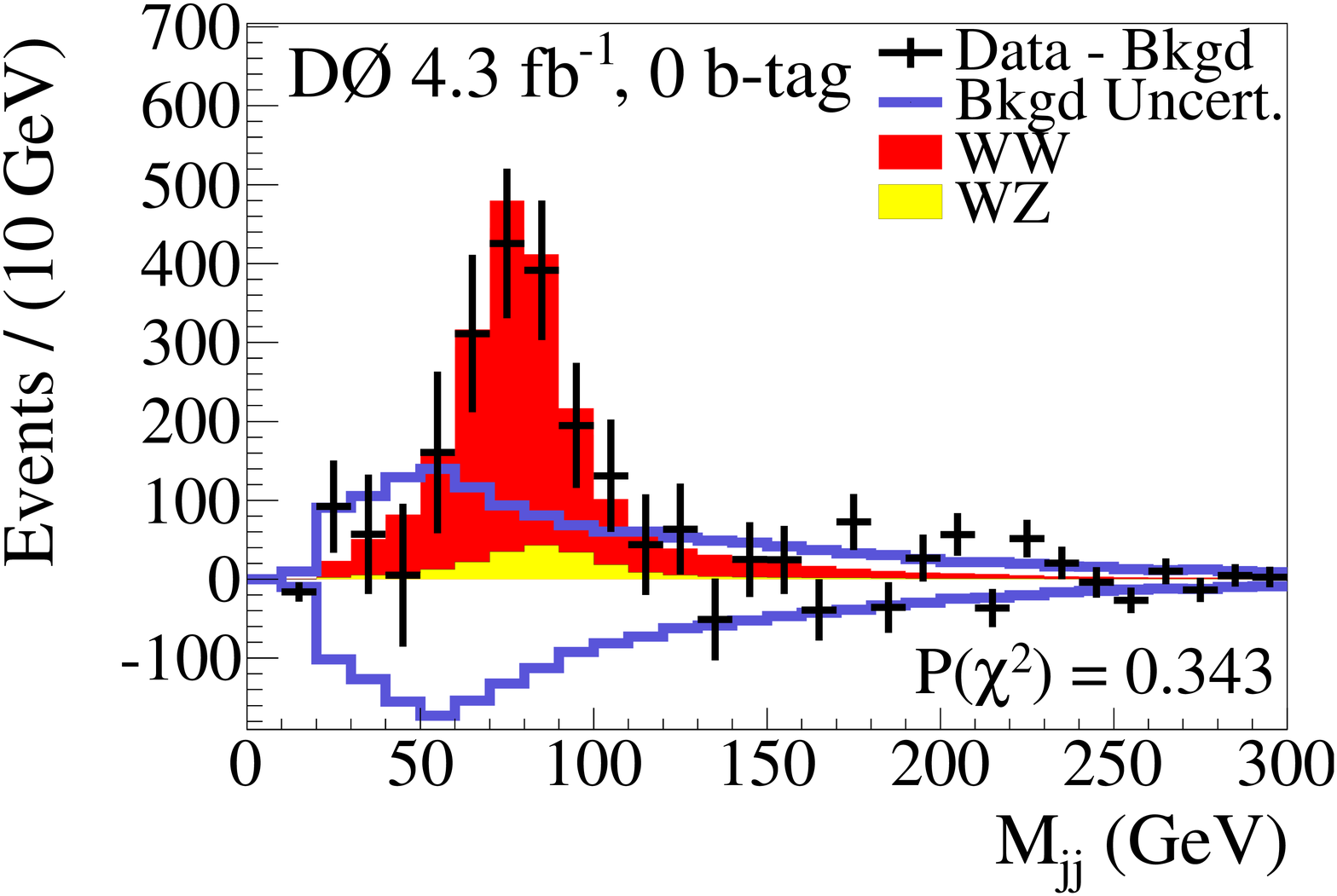}
      \includegraphics[width=2.3in]{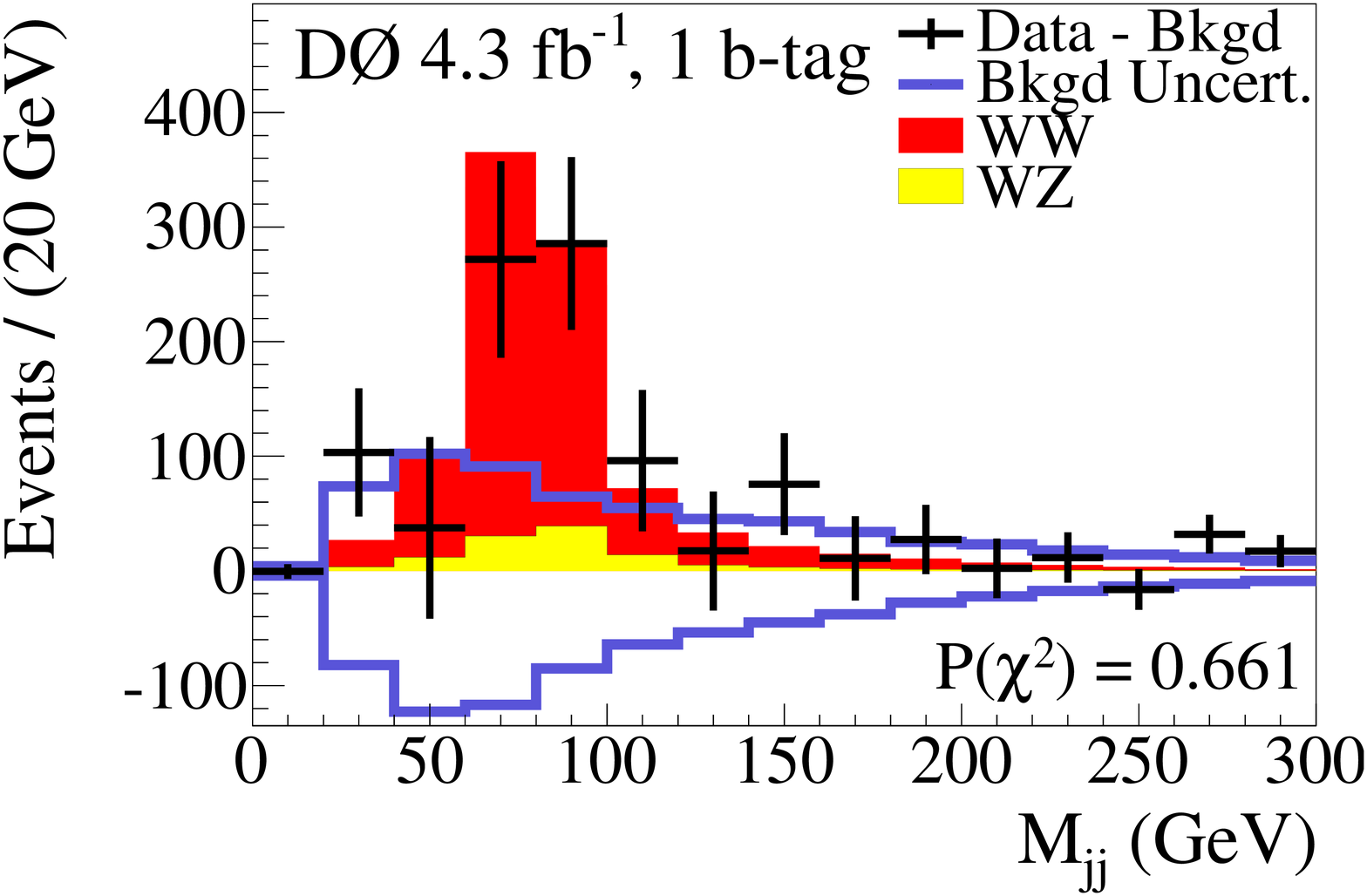}
      \includegraphics[width=2.3in]{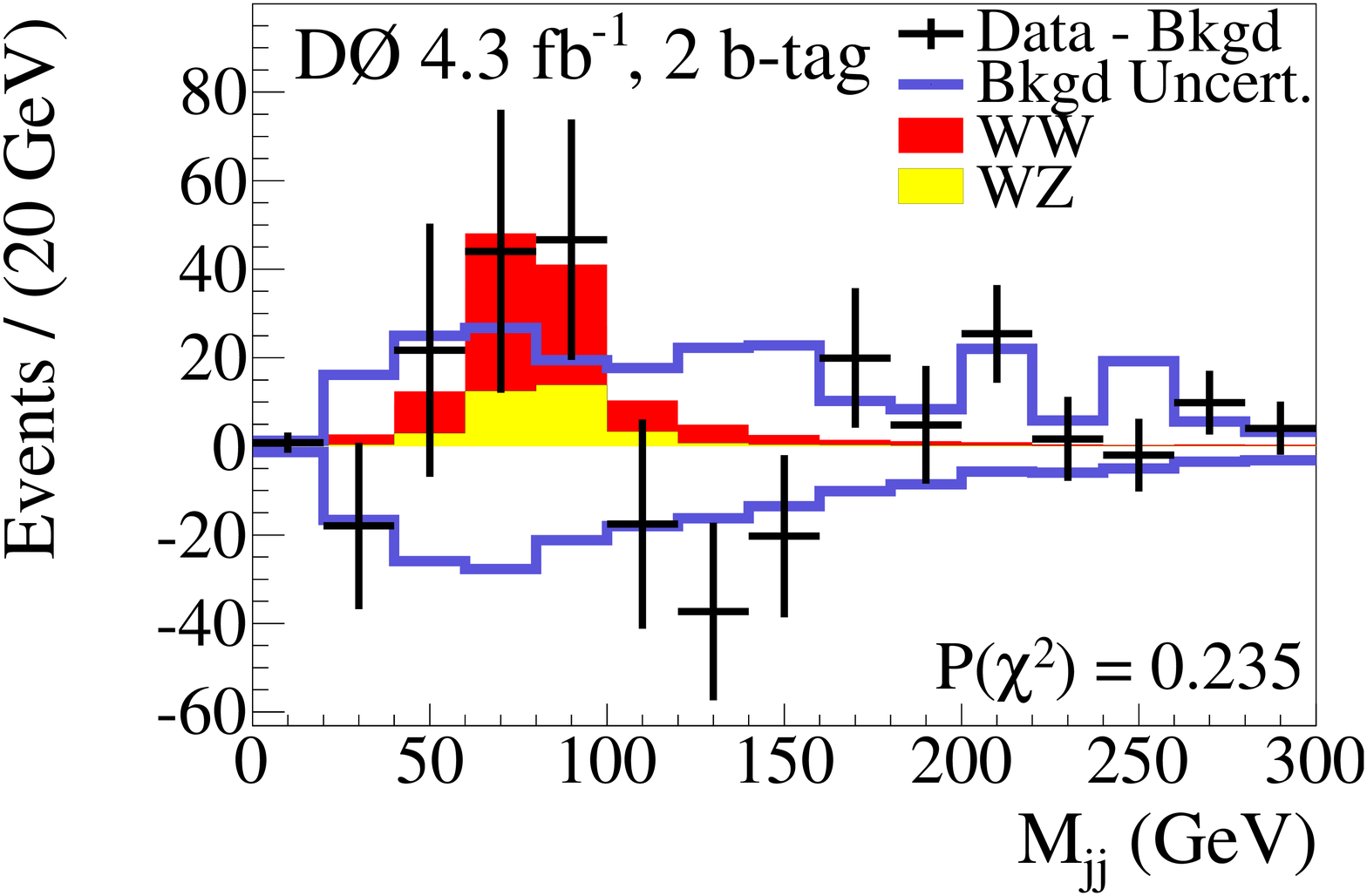}

      \caption{(color online) A comparison of the measured $WW$ and
	$WZ$ signals (filled histograms) to background-subtracted data
	(points) in the dijet mass distribution (summed over electron
	and muon channels) for 0, 1, and 2-tag sub-channels after
	the combined fit to data using the dijet mass distribution.
	Also shown is the $\pm$1 standard deviation uncertainty on the
	background prediction.  The $\chi^{2}$ fit probability, 
        P$(\chi^{2})$, is based on the residuals using data and MC 
        statistical uncertainties.}
      
      \label{fig:Fig4}
    \end{centering}
  \end{figure*}

  The fit of the total $WV$ cross section using the RF output 
  distributions yields $\sigma(WV)= 19.6~^{+3.2}_{-3.0}$~pb, 
  corresponding to an observed (expected) significance of 7.9 (5.9) 
  standard deviations (s.d.).  Figure~\ref{fig:Fig1} shows the 
  background-subtracted RF output distribution summed over all 
  sub-channels after the fit.  As a cross check, we perform the 
  measurement using the dijet mass distribution in place of the 
  full RF output distribution~\cite{bib:EPAPS}.  This measurement 
  yields a $WV$ cross section of $\sigma(WV)= 18.3~^{+3.8}_{-3.6}$~pb, 
  consistent with that obtained using the RF output distribution.
  
  The fit is then performed with the signal divided into the separate
  $WW$ and $WZ$ components, which are allowed to float independently.
  The result of this simultaneous fit of $\sigma(WW)$ and $\sigma(WZ)$
  using the RF output distributions is shown in Fig.~\ref{fig:Fig2}.
  It yields $\sigma(WW)=
  15.9~^{+1.9}_{-1.5}$~(stat)$~^{+3.2}_{-2.9}$~(syst)~pb and $\sigma(WZ)=
  3.3~^{+3.4}_{-2.7}$~(stat)$~^{+2.2}_{-1.8}$~(syst)~pb.  The RF output 
  distributions for the 0, 1, and 2-tag sub-channels from this fit are 
  shown in Fig.~\ref{fig:Fig3}.  This measurement is also verified fitting 
  the dijet mass distribution, which yields
  $\sigma(WW)=13.3~^{+2.8}_{-2.2}$~(stat)$~^{+3.6}_{-2.9}$~(syst)~pb and 
  $\sigma(WZ)=5.4~^{+2.7}_{-2.6}$~(stat)$~^{+4.5}_{-4.3}$~(syst)~pb.
  Figure~\ref{fig:Fig4} shows plots for the background-subtracted 
  dijet mass after the dijet mass fit.

  We also perform a fit in which we constrain the $WW$ cross section
  to its SM prediction with a Gaussian prior equal to the theoretical
  uncertainty of 7\%~\cite{bib:Campbell}.  The fit of the RF output
  distribution yields a $WZ$ cross section of $\sigma(WZ)=$ 6.5 $\pm$
  0.9~(stat) $\pm$ 3.0~(syst)~pb with observed (expected) significance 
  of 2.2 (1.2)~s.d., and the dijet mass fit yields 
  $\sigma(WZ)=$ 6.7 $\pm$ 1.0~(stat) $\pm$ 3.9~(syst)~pb with observed 
  (expected) significance of 1.7 (0.9)~s.d.  As expected, now that 
  $\sigma(WW)$ is constrained to the SM prediction, the fit requires a 
  higher rate for $WZ$ in order to account for the excess of signal-like 
  events.

  In summary, we have measured the cross section for total $WV$ 
  production to be $\sigma(WV)= 19.6~^{+3.2}_{-3.0}$~pb ($V=W$ or $Z$) 
  with a significance of \wvRFsd~s.d.~above the background-only 
  hypothesis.  This result demonstrates the ability of the D0 
  experiment to measure a dijet signal in a background-dominated 
  final state directly relevant to low mass Higgs searches.  
  Furthermore, we have used $b$-jet tagging to measure the 
  contributions from $WW$ and $WZ$ and measured the cross sections 
  for the separate processes to be $\sigma(WW)=15.9~^{+3.7}_{-3.2}$~pb 
  and $\sigma(WZ)=\wzRF~^{+\wzRFeu}_{-\wzRFed}$~pb.  Although we 
  cannot yet claim 3~s.d.~evidence of a $WZ$ signal in the $\ell\nu{jj}$ 
  final states, the extracted $WV$ and $WZ$ cross sections are in 
  agreement with the SM prediction and their precise measurement  
  represents an independent test to new physics which could manifest 
  itself differently in different final states.

%
We thank the staffs at Fermilab and collaborating institutions,
and acknowledge support from the
DOE and NSF (USA);
CEA and CNRS/IN2P3 (France);
FASI, Rosatom and RFBR (Russia);
CNPq, FAPERJ, FAPESP and FUNDUNESP (Brazil);
DAE and DST (India);
Colciencias (Colombia);
CONACyT (Mexico);
KRF and KOSEF (Korea);
CONICET and UBACyT (Argentina);
FOM (The Netherlands);
STFC and the Royal Society (United Kingdom);
MSMT and GACR (Czech Republic);
CRC Program and NSERC (Canada);
BMBF and DFG (Germany);
SFI (Ireland);
The Swedish Research Council (Sweden);
and
CAS and CNSF (China).

\clearpage

\section{Appendix}

\section{Input Variables to the Random Forest Classifier}
Here we define the fifteen variables used as inputs to the RF
classifier.  The observed distribution for each variable is shown in
Figs.~\ref{fig:btags},~\ref{fig:rf_inputs1} and~\ref{fig:rf_inputs2} 
along with the predicted distribution after the fit of the total $WV$ 
($V=W,Z$) cross section using the RF output distribution.

The RF input variables can be classified into three
categories: (i) $b$-jet identification variables, (ii) kinematics of
individual final state particles, and (iii) kinematics of multiple
final state particles.  Several variables are calculated using the
four-momentum of the dijet system, which we define as the sum of the
four-momenta of the two highest $p_T$ jets.  We also reconstruct a
$W\to\ell\nu$ candidate $W^{\ell\nu}$, from the charged lepton and 
the \met.  The neutrino from the $W$$\to
\ell \nu$ decay is assigned the transverse momentum defined by \met\
and a longitudinal momentum that is calculated assuming the mass of
the $\ell \nu$ system is 80.4~GeV.  Of the two possible solutions, we
choose the real component that provides the smaller total invariant
mass of all objects in the event.

\begin{itemize}

  \item \textbf{$b$-jet Identification Variables:}
  
  The NN $b$-tagger has 12 operating points characterized by different 
  purities.  Each jet is assigned an integer $b$-tag value based on the 
  highest purity operating point that it passes.
  
  \begin{enumerate}
  \item Max $b$-tag Value: The greater $b$-tag value of the two
  highest $p_T$ jets.  The neural network $b$-tagger has 12 operating
  points of increasing purity and the $b$-tag value corresponds to the
  highest operating point satisfied by the jet (or zero if the jet did
  not satisfy any of the operating points).
  
  \item Min $b$-tag Value: The lesser $b$-tag value of the two highest
  $p_T$ jets.
  \end{enumerate}

  These variables are shown in Fig.~\ref{fig:btags}, both in logarithmic 
  and linear scales. 

  \item \textbf{Kinematics of Individual Final State Particles:}
  \begin{enumerate}
   
  \item $p_T(\ell)$: The $p_T$ of the charged lepton.
   
  \item $p_T(\mathrm{jet}_1)$: The highest jet $p_T$.

  \item $p_T(\mathrm{jet}_2)$: The second highest jet $p_T$.

  \item \met: The imbalance in transverse energy determined from
  the energy measured in each calorimeter cell and then corrected 
  for reconstructed muons, jets, and electrons/photons.
  \end{enumerate}

  \item \textbf{Kinematics of Multiple Final State Particles:}
  \begin{enumerate}
  
  \item $M_{jj}$: The invariant mass of the dijet system reconstructed 
  from the two highest $p_{T}$ jets.

  \item $M_T^{\ell\nu} =
  \sqrt{2\;p_T^{\ell}\;\met\;(1-\cos(\Delta\phi(\ell,\met)))}$: The
  transverse $W$ mass reconstructed from the charged lepton and
  the \met.

  \item $H_T = p_T(\mathrm{jet}_1) + p_T(\mathrm{jet}_2)$: The
  scalar sum of the two highest jet $p_T$s.

  \item
  $p_{T}^\mathrm{rel}(\mathrm{dijet},\mathrm{jet}_{1})^{W} =
  {|\vec{p}_T(\mathrm{jet}_1)\times\hat{p}_T(\mathrm{jet}_1+\mathrm{jet}_2)|}$:
  The magnitude of the leading jet transverse momentum perpendicular to
  the dijet system in the rest frame of the $W^{\ell\nu}$
  candidate;

  \item $p_{T}^\mathrm{rel}(\mathrm{dijet},\mathrm{jet}_{2}) =
  {|\vec{p}_T(\mathrm{jet}_2)\times\hat{p}_T(\mathrm{jet}_1+\mathrm{jet}_2)|}$:
  The magnitude of the second-leading jet transverse momentum perpendicular to
  the dijet system in the laboratory frame.
  
  \item $k_{T}^\mathrm{min} = \Delta
  R(\mathrm{jet}_1,\mathrm{jet}_2)\frac{p_T(\mathrm{jet}_2)}{p_T(\ell)+\met}$:
  The angular separation between the two jets of highest $p_T$,
  weighted by the ratio of the transverse momentum of the
  second-leading jet and a scalar sum of the transverse momenta of the $W^{\ell\nu}$ 
  constituents.
  
  \item $\cos(\angle(\mathrm{dijet},\mathrm{jet}_1))$: The cosine of
  the angle between the momentum vectors of the dijet system and the
  highest $p_T$ jet in the laboratory frame.

  \item $\cos(\angle(W^{\ell\nu},\mathrm{jet}_1))^{jj}$: Cosine of the
  angle between the momentum vectors of the leading jet and the
  $W^{\ell\nu}$ candidate, evaluated in the rest frame of the dijet
  system.

  \item Centrality: The scalar sum of transverse momenta of the charged 
  lepton and all jets in the event divided by the sum of their energies. 

  \end{enumerate}
\end{itemize}

  \begin{figure*}[hbt]
    \begin{centering}
      \includegraphics[width=3.25in]{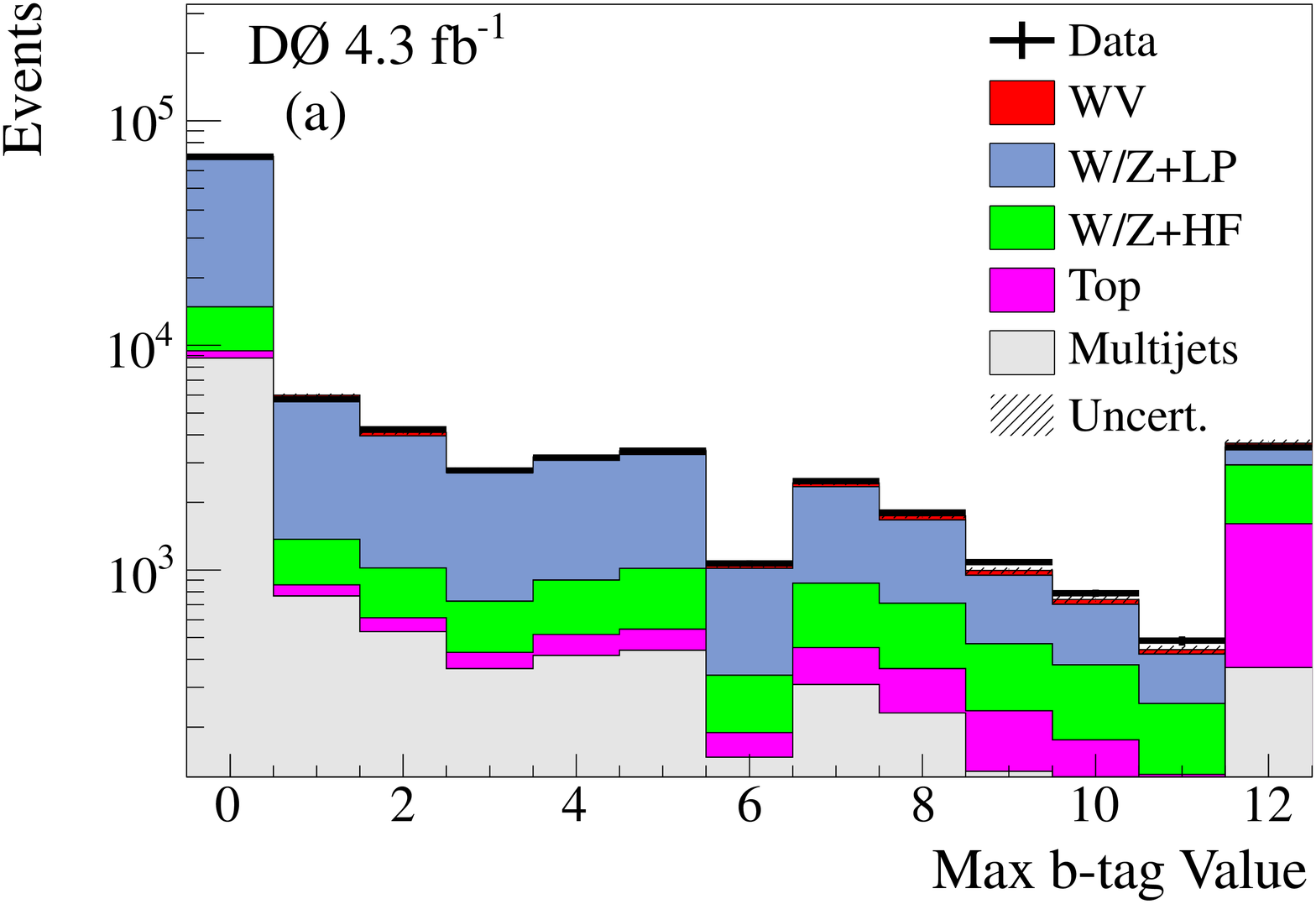}
      \includegraphics[width=3.25in]{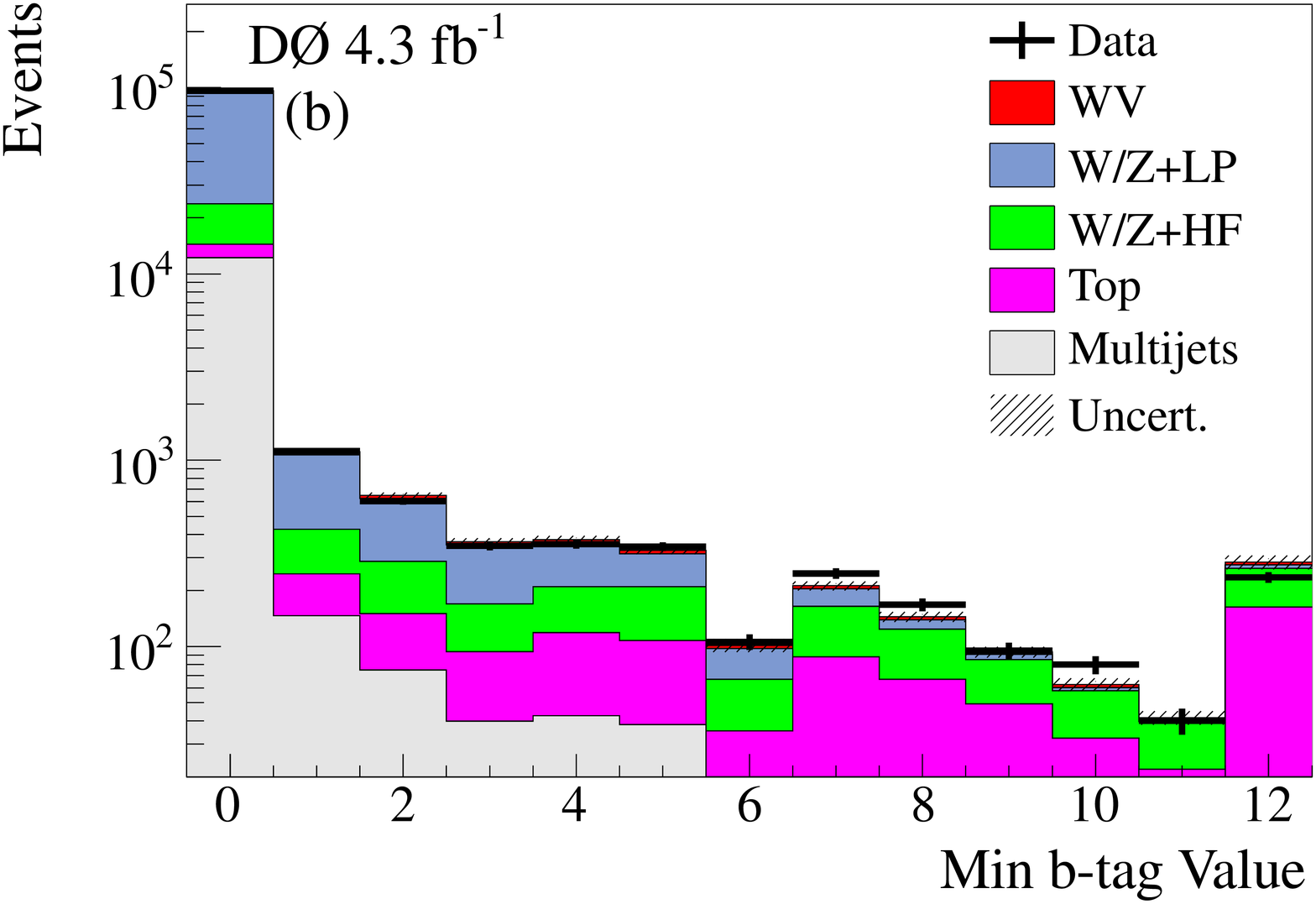} \\
      \includegraphics[width=3.25in]{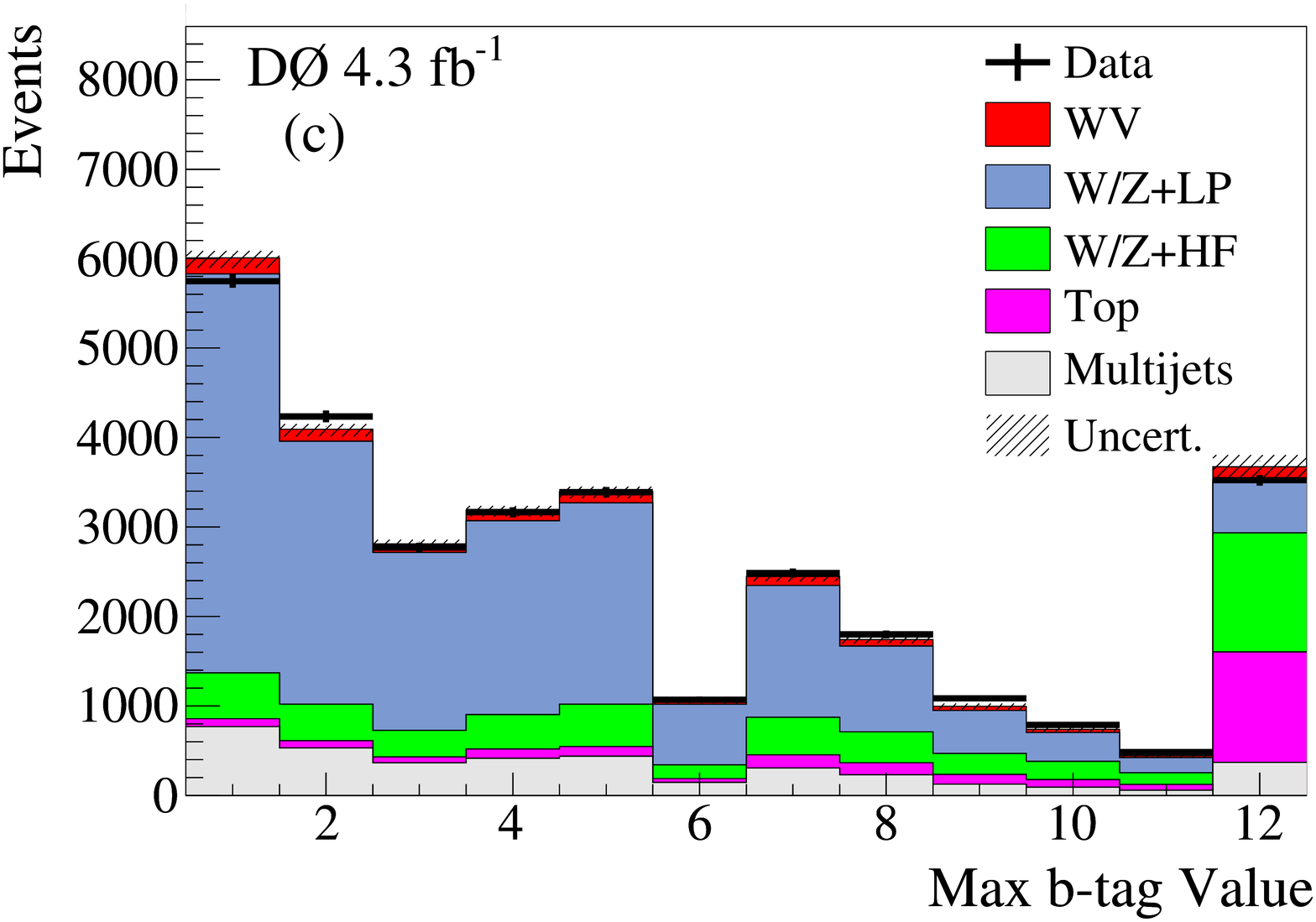}
      \includegraphics[width=3.25in]{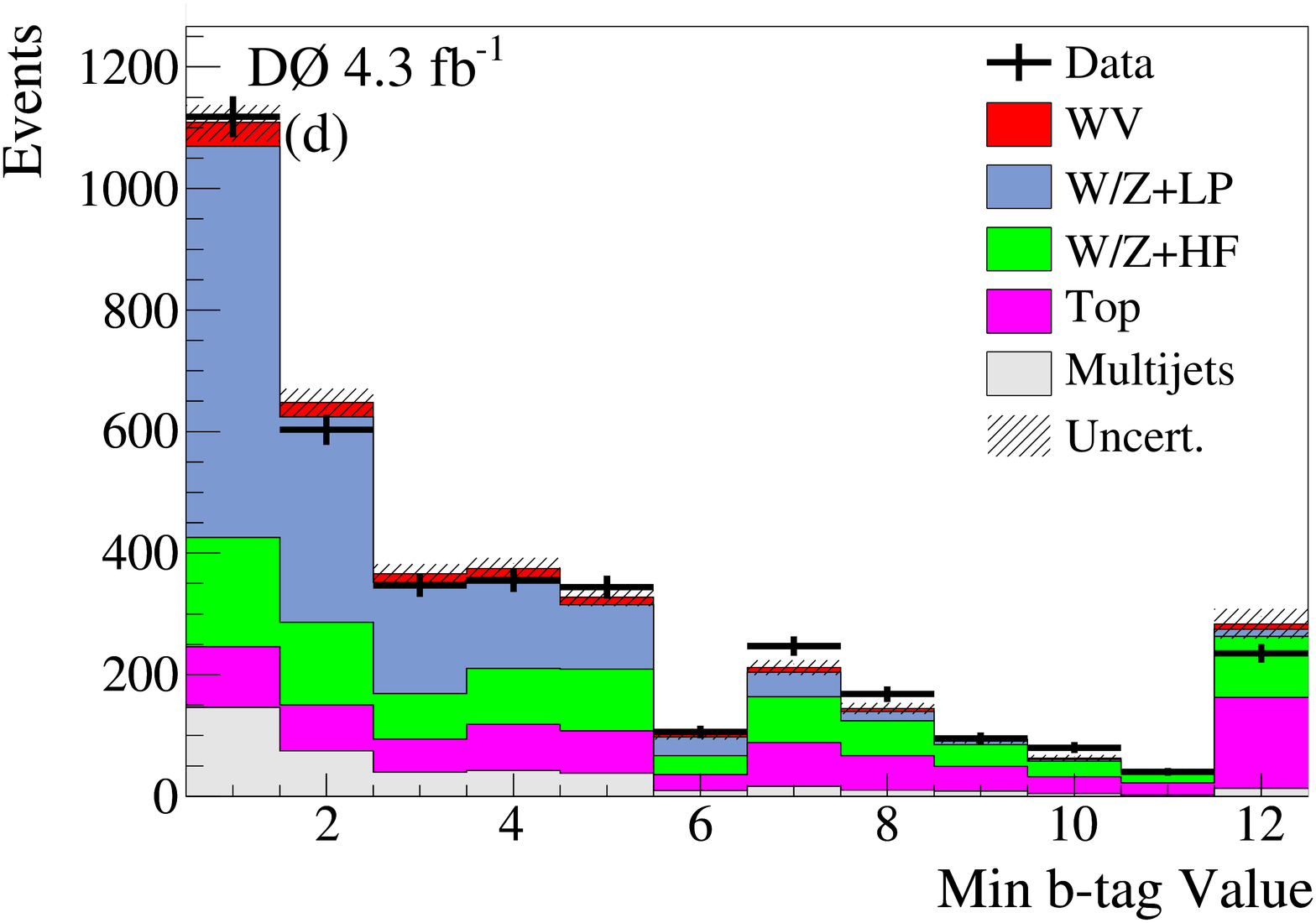} \\
    \caption{(color online) Distributions of the $b$-jet identification 
        variables used as inputs to the RF classifier (first two of fifteen)
	for electron and muon channels combined, with logarithmic ((a) and (b)) 
	and linear ((c) and (d)) scales.  To better show the $WW$ and $WZ$ signals 
	the lowest bin is cut off in the distributions with a linear scale.
	The signal and background predictions and 
	the systematic uncertainty band are evaluated after the fit of the 
	total $WV$ cross section in the RF output distribution.  Definitions 
	for each variable are provided in the text (LP denotes light partons 
	such as $u$, $d$, $s$ or gluon, and HF denotes heavy-flavor such as 
	$c\bar{c}$ or $b\bar{b}$).} \label{fig:btags}
    \end{centering}
  \end{figure*}
  \begin{figure*}[hbt]
    \begin{centering}
      \includegraphics[width=3.25in]{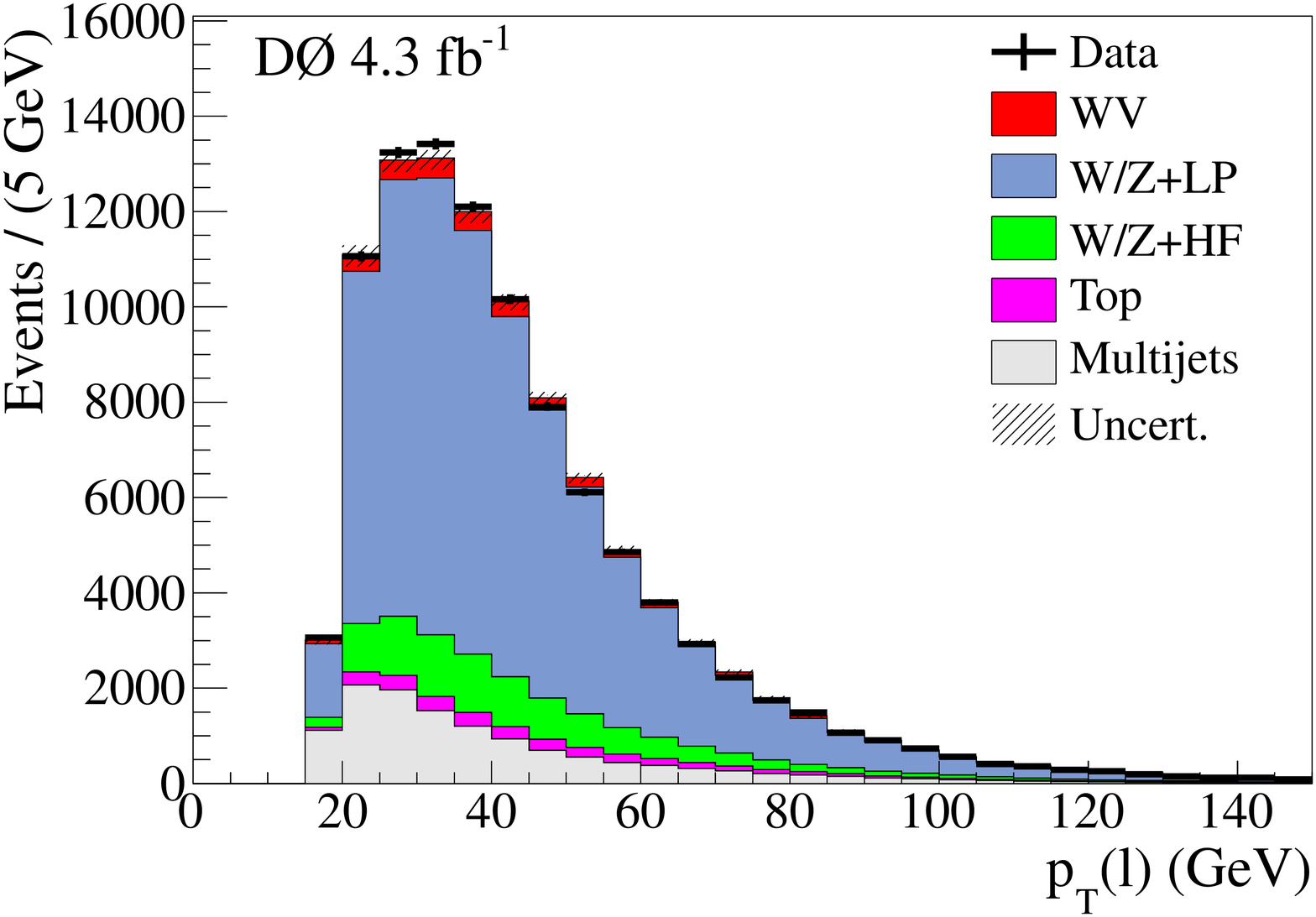}
      \includegraphics[width=3.25in]{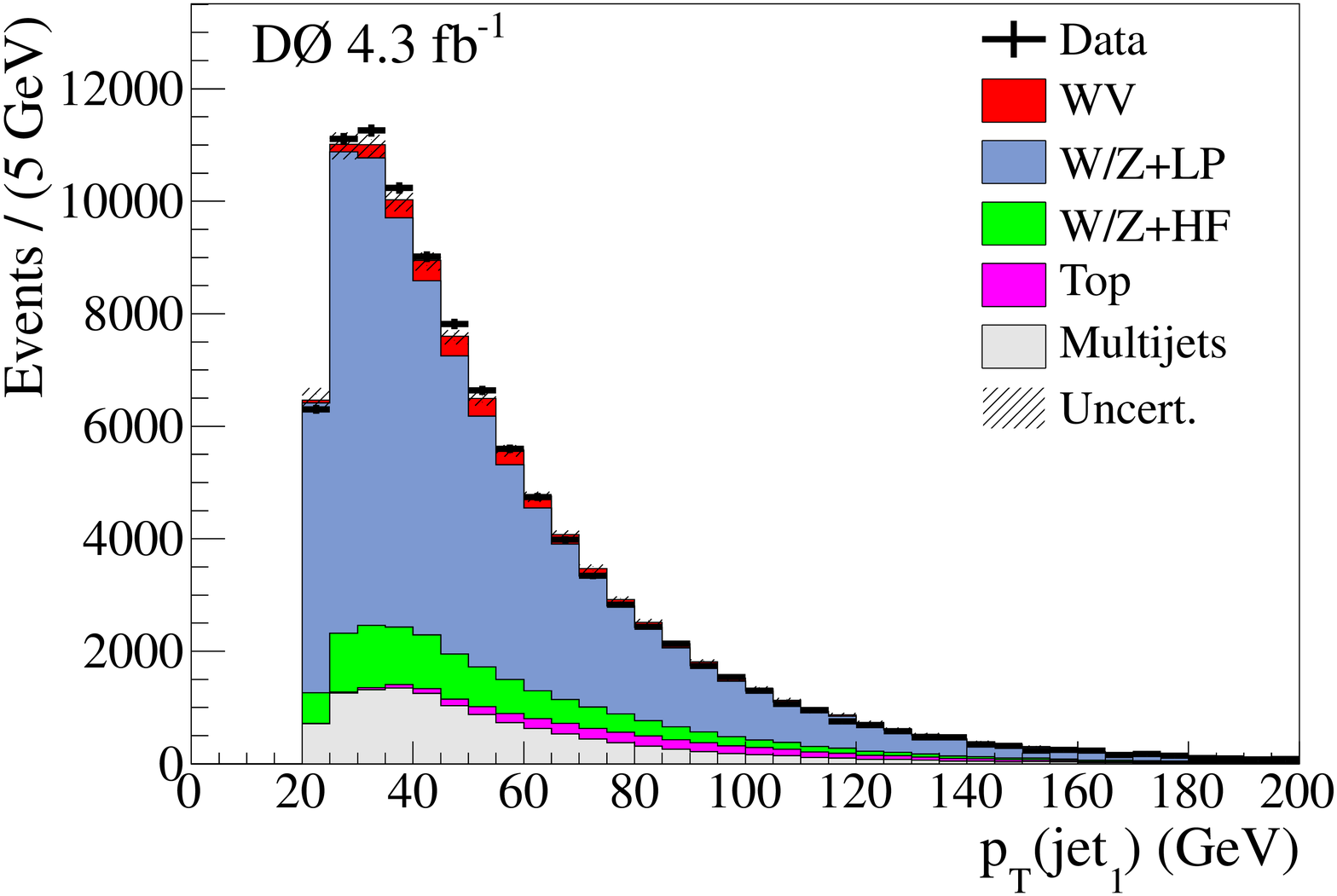} \\
      \includegraphics[width=3.25in]{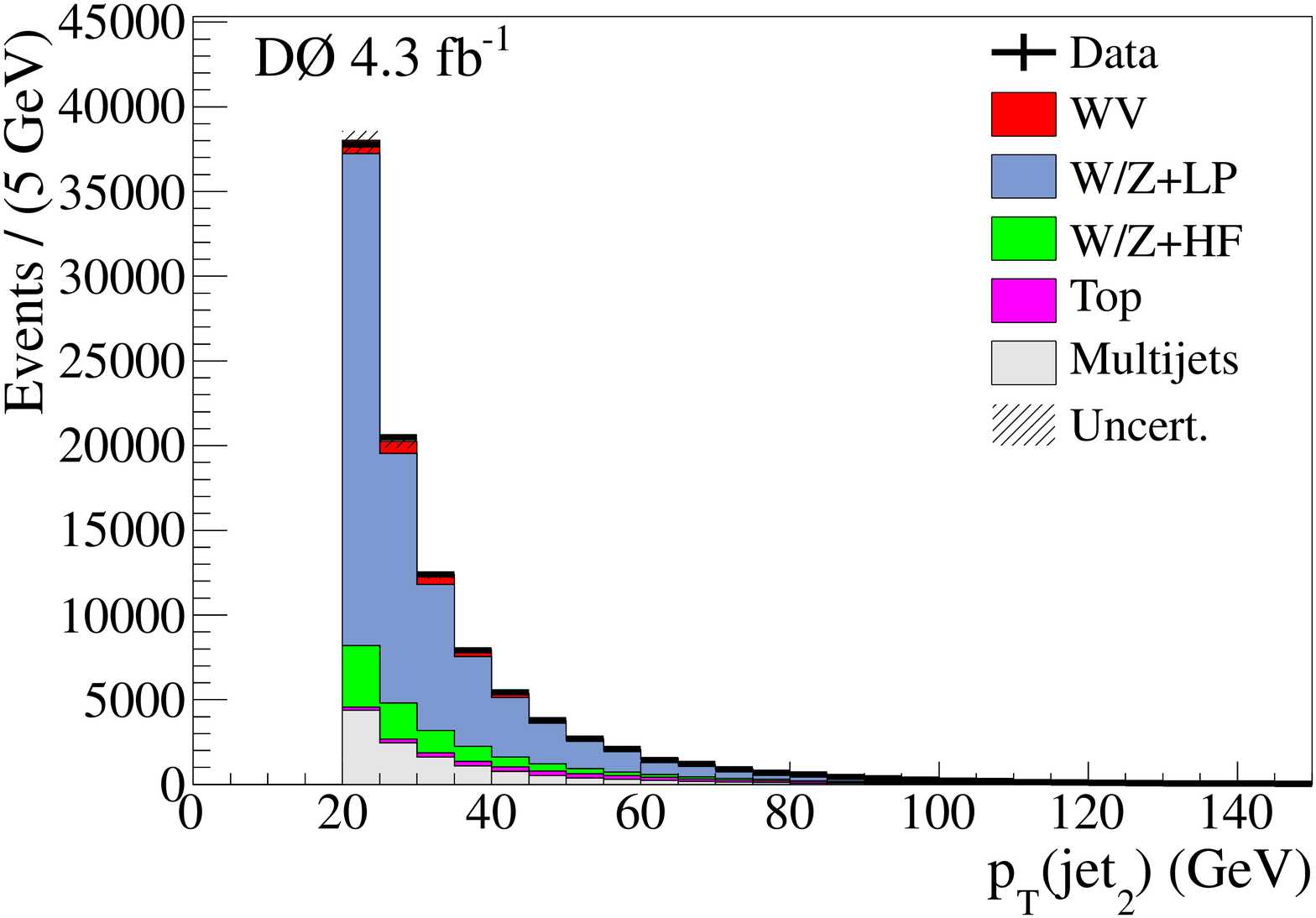}
      \includegraphics[width=3.25in]{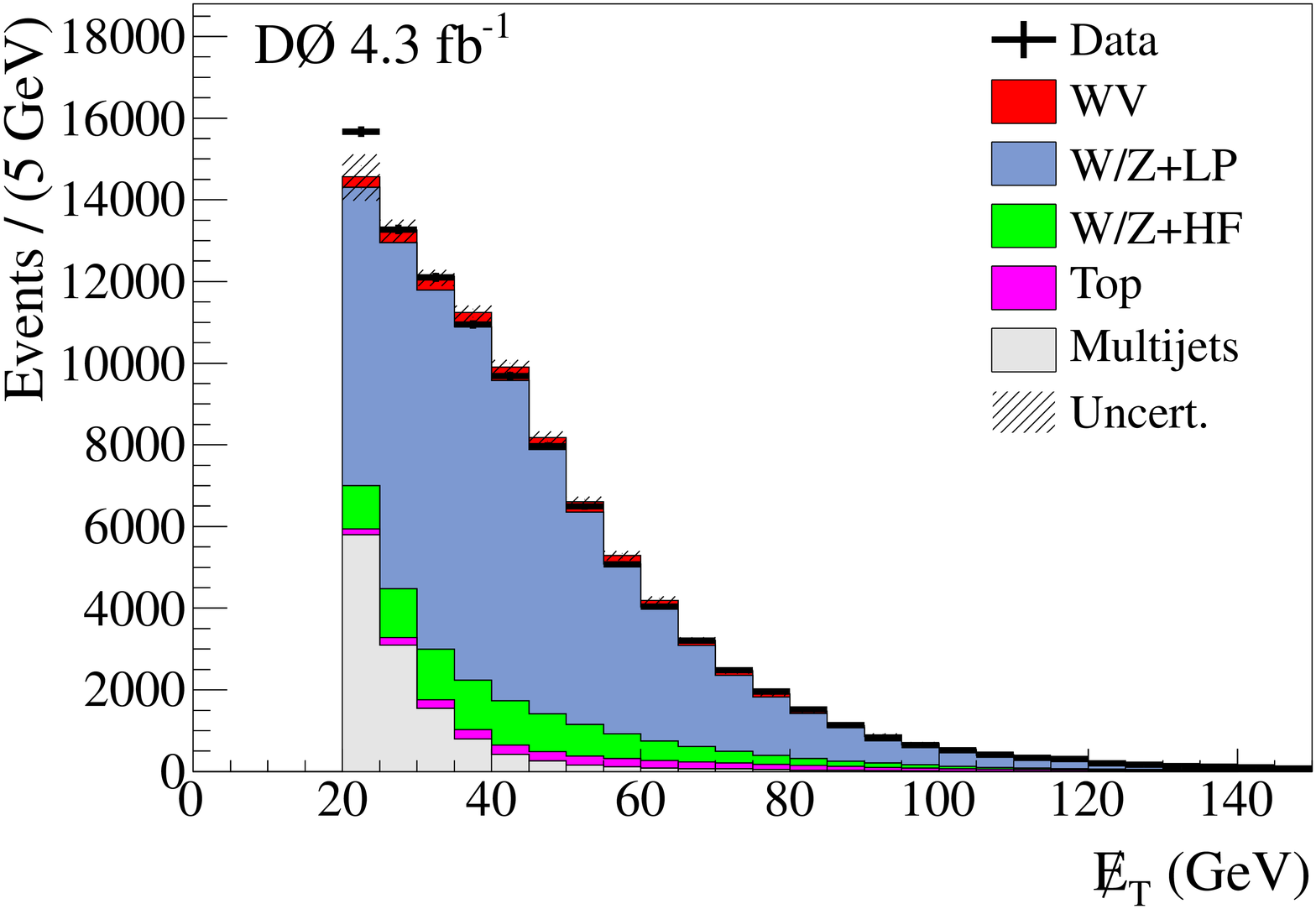} \\
      \includegraphics[width=3.25in]{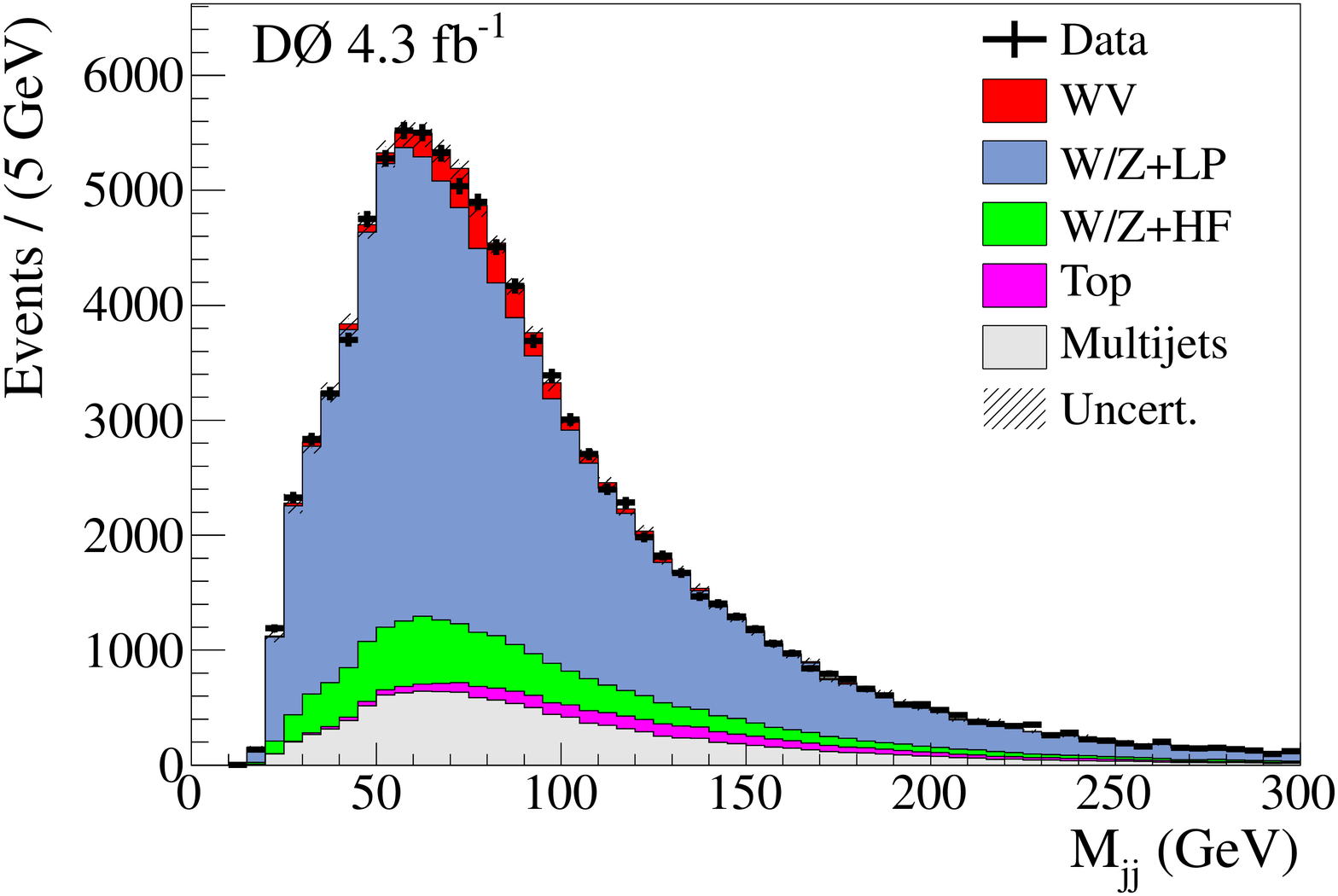}
      \includegraphics[width=3.25in]{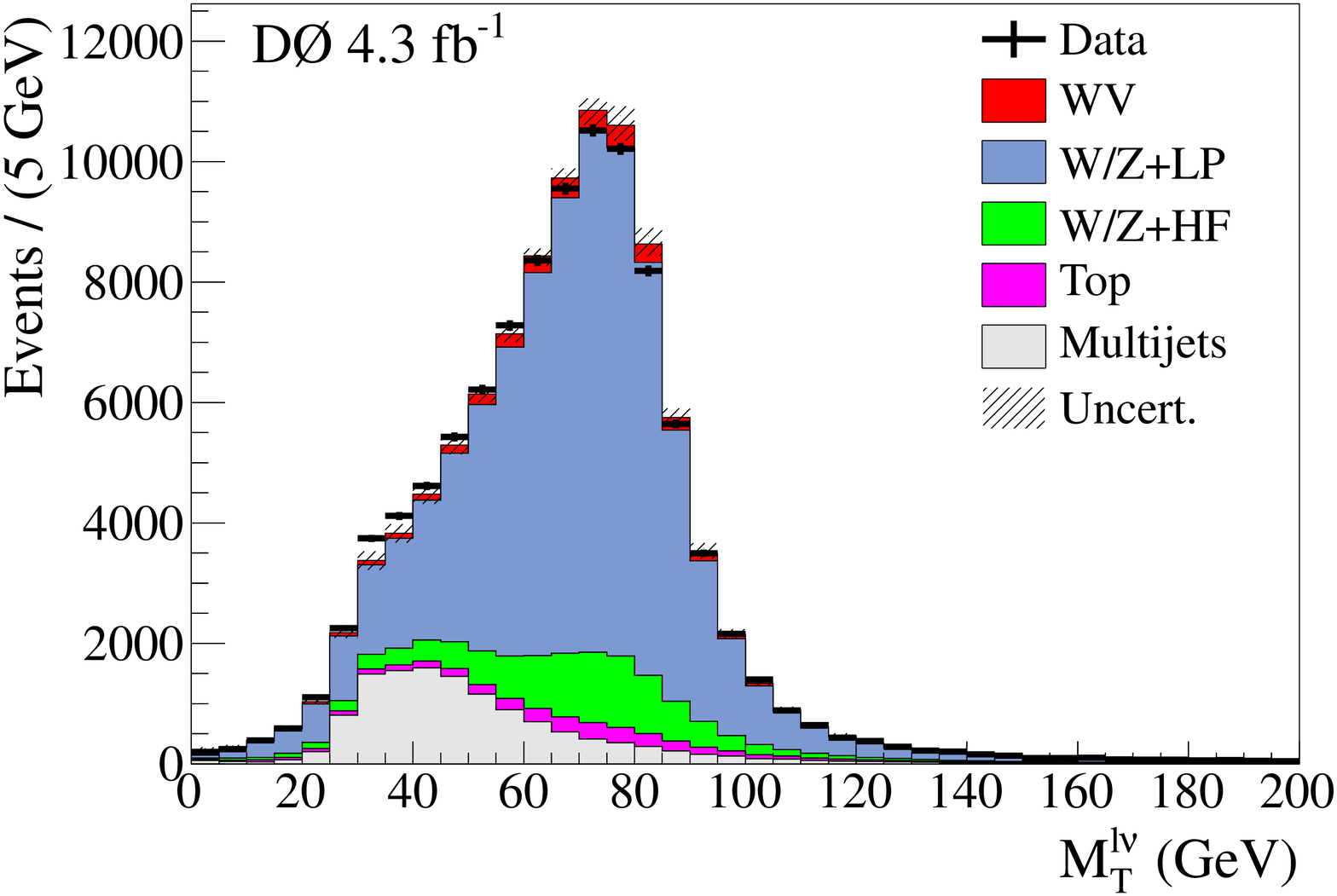} \\
      \includegraphics[width=3.25in]{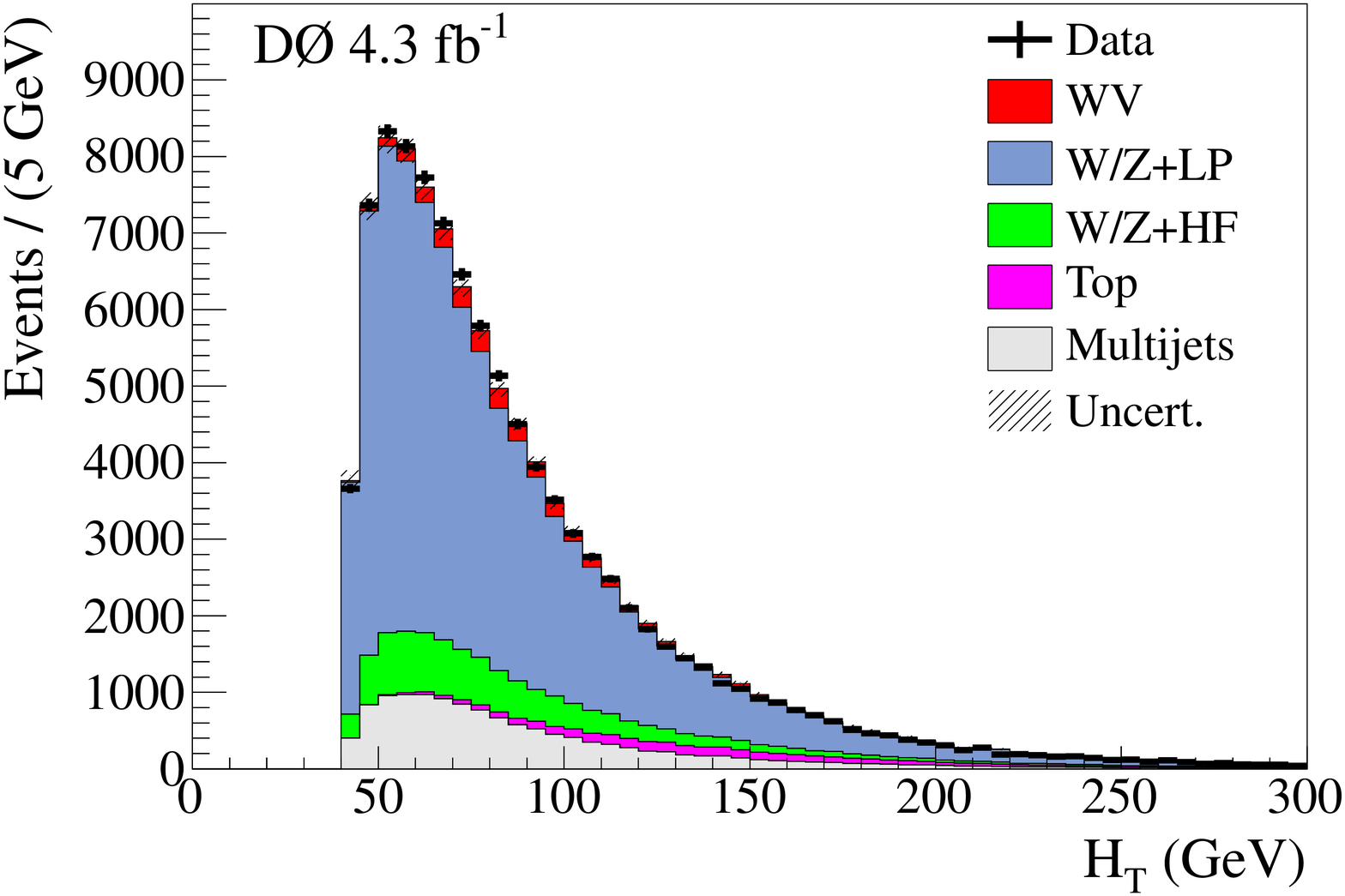}
      \includegraphics[width=3.25in]{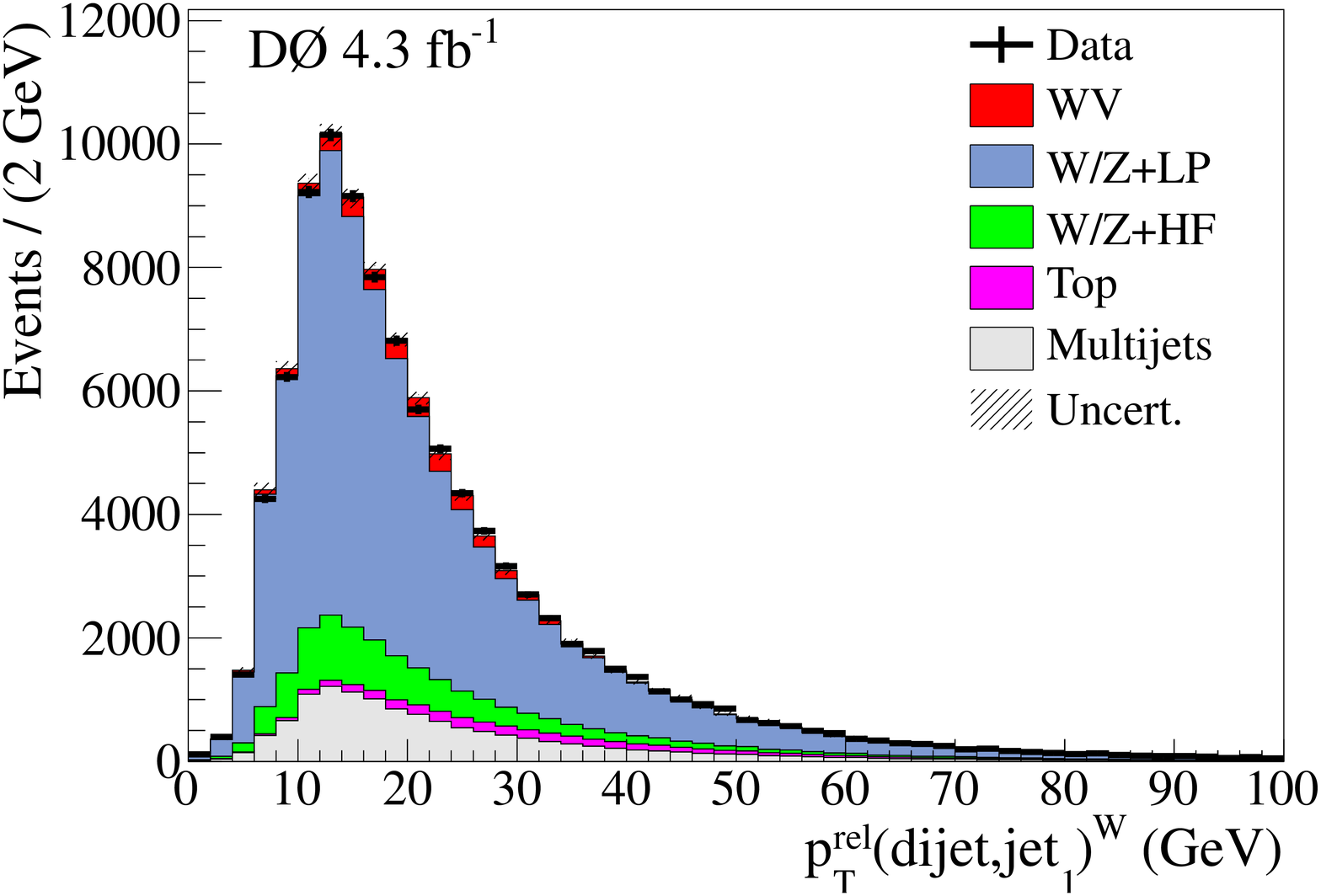} \\
      \caption{(color online) Distributions of the variables (next
	eight of fifteen) used as inputs to the RF classifier for
	electron and muon channels combined, and before $b$-tagging.
	The signal and background predictions and the systematic
	uncertainty band are evaluated after the fit of the total $WV$
	cross section in the RF output distribution.  Definitions for
	each variable are provided in the text 
        (LP denotes light partons such as $u$, $d$, $s$ 
        or gluon, and HF denotes heavy-flavor such as $c\bar{c}$ 
        or $b\bar{b}$).}
	\label{fig:rf_inputs1}
    \end{centering}
  \end{figure*}

  \begin{figure*}[hbt]
    \begin{centering}
      \includegraphics[width=3.25in]{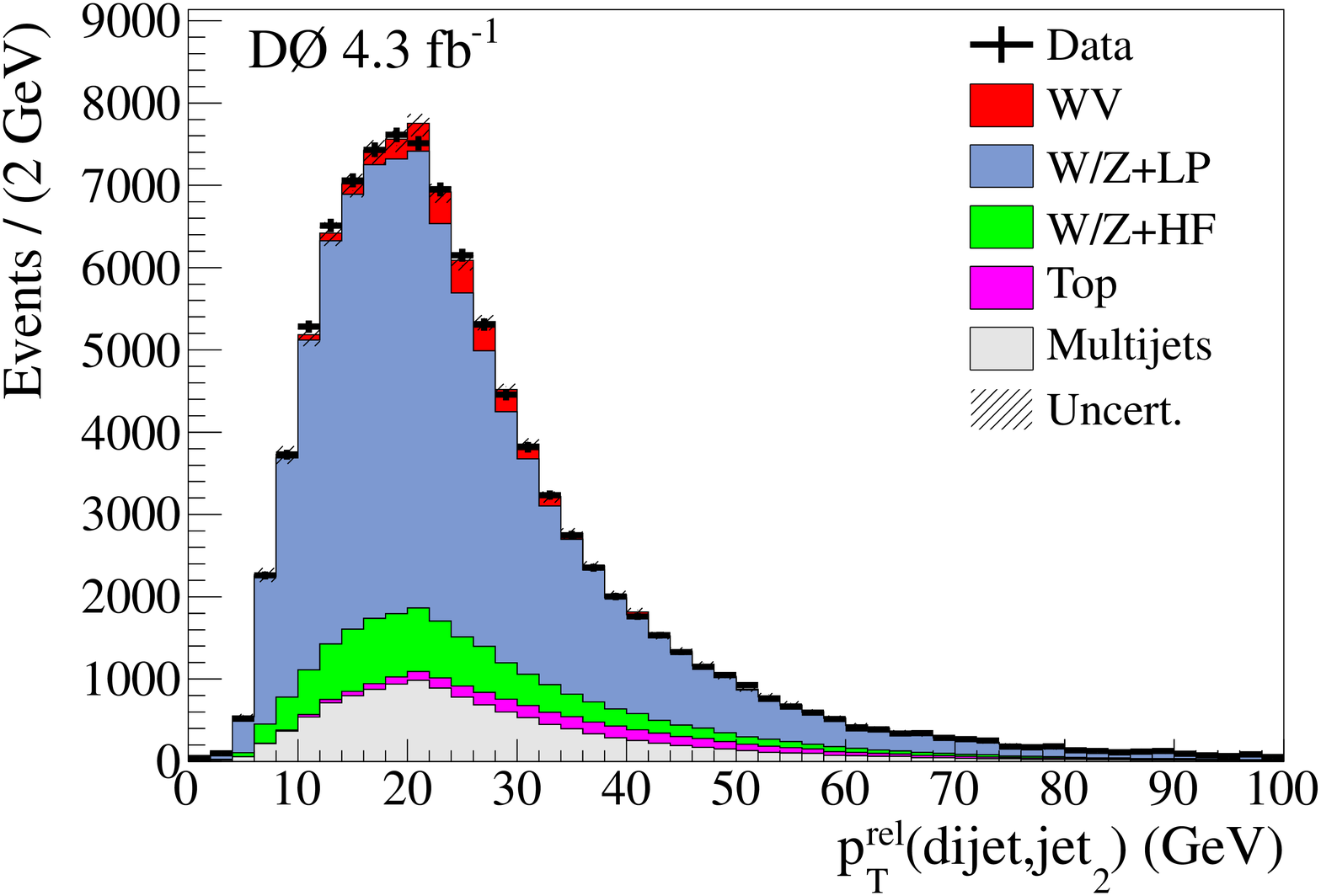}
      \includegraphics[width=3.25in]{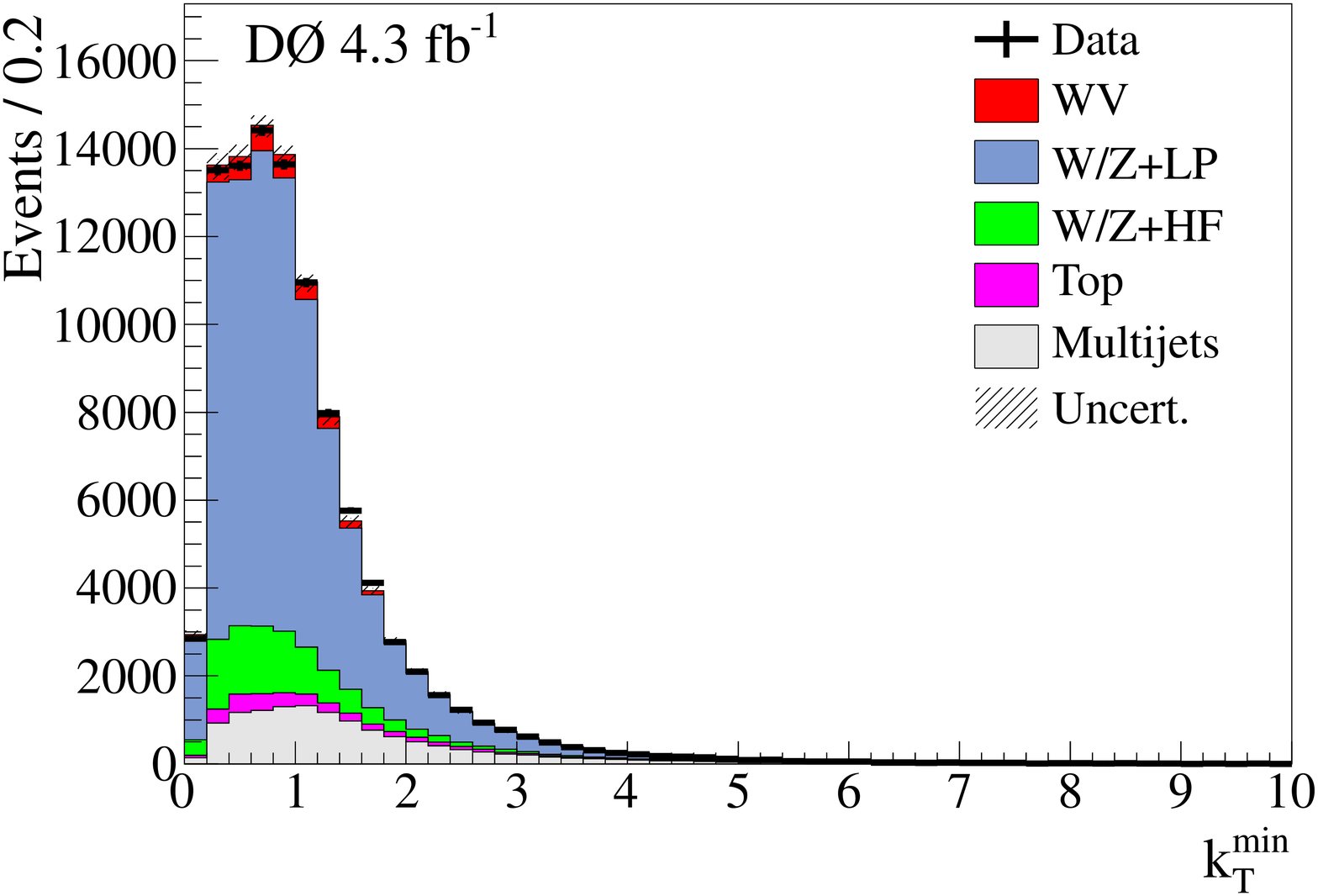} \\
      \includegraphics[width=3.25in]{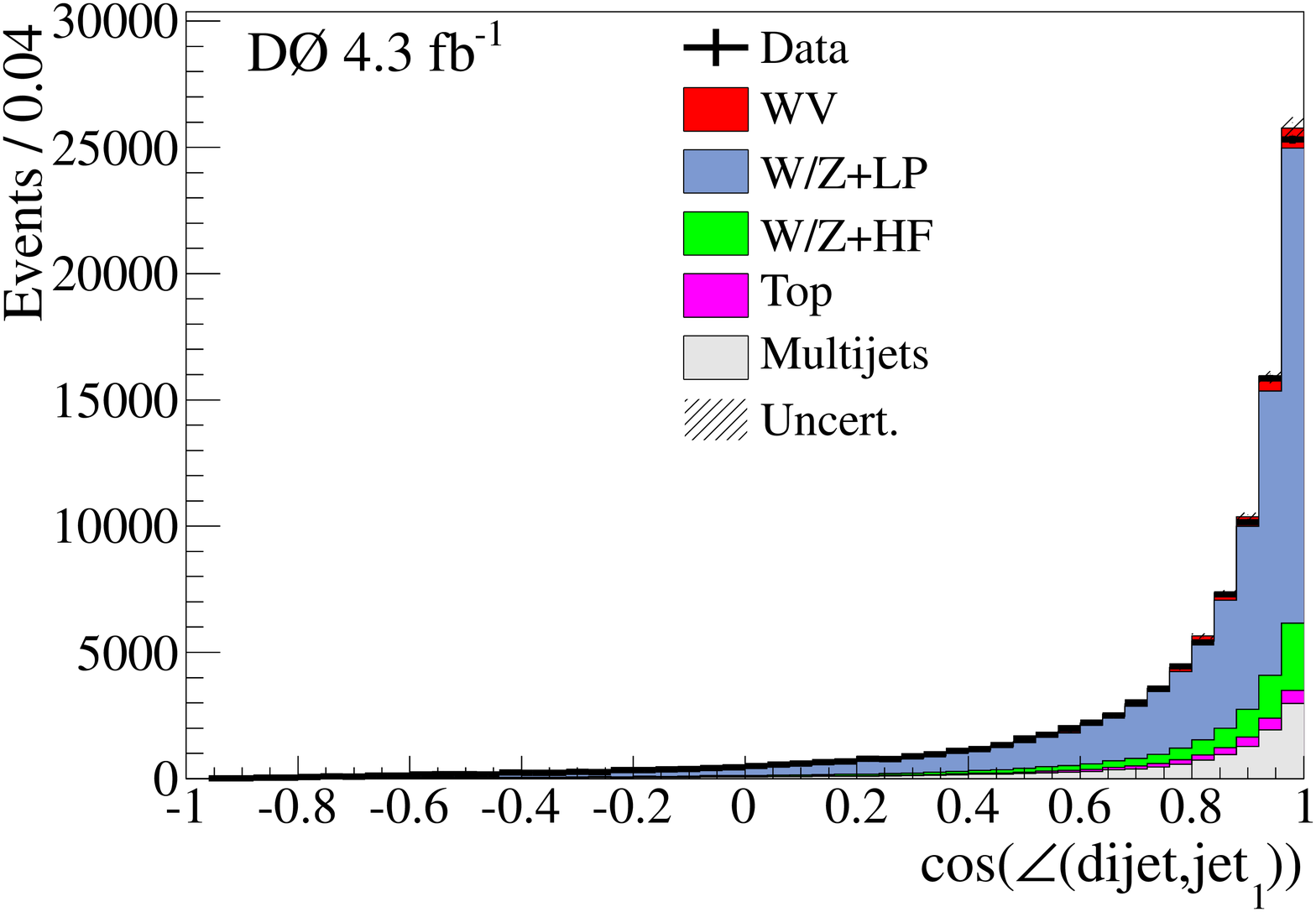}
      \includegraphics[width=3.25in]{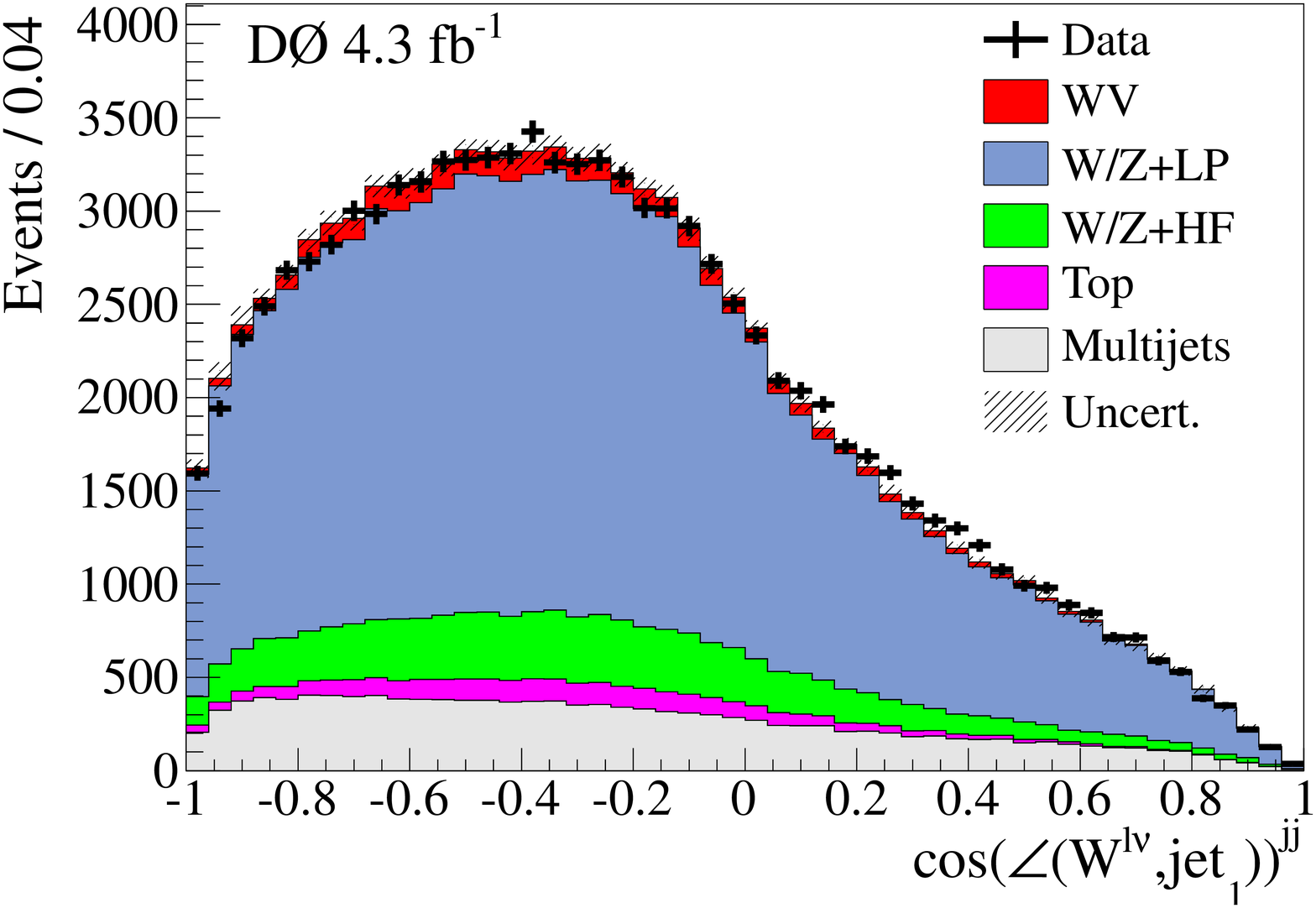} \\
      \includegraphics[width=3.25in]{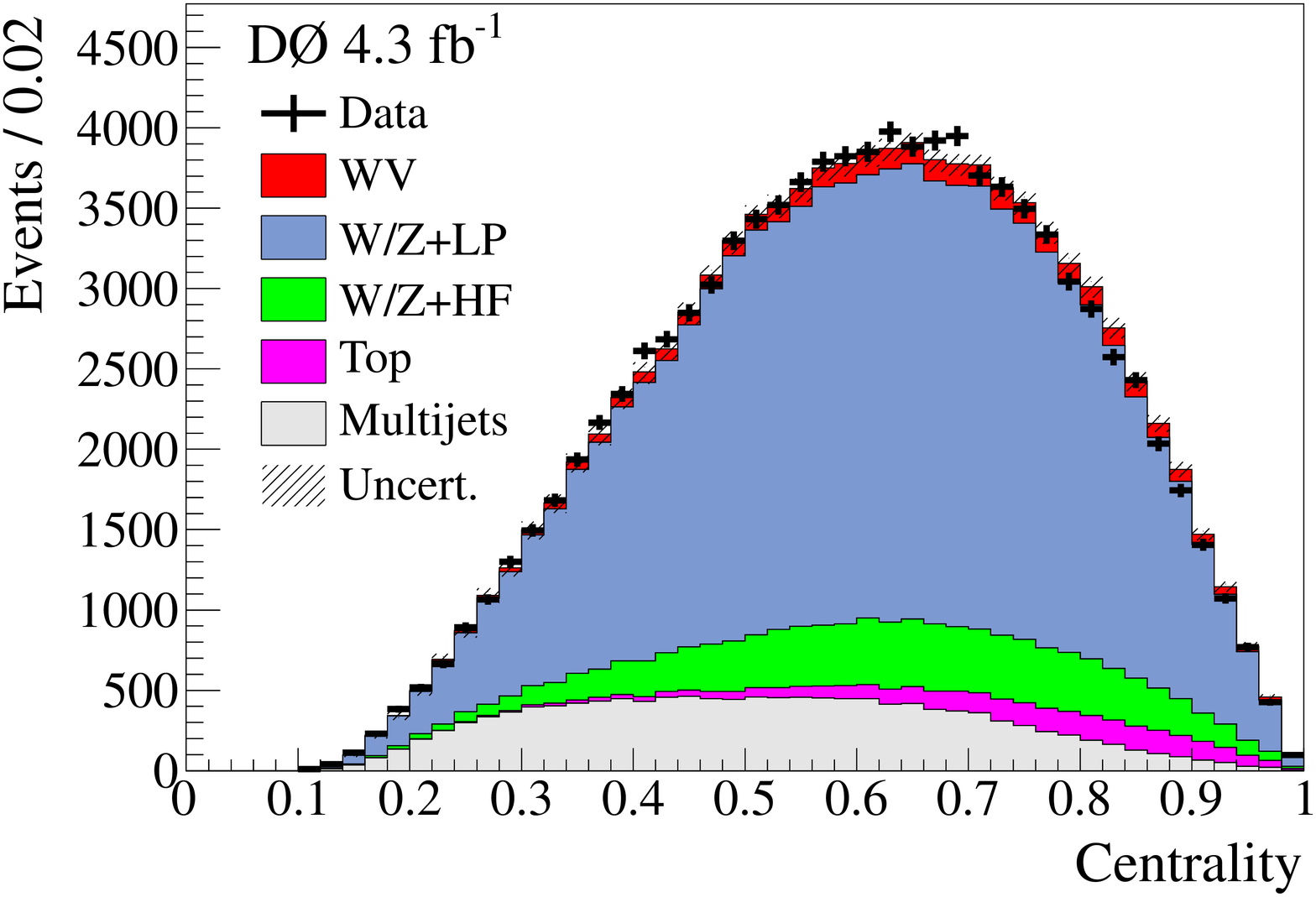}
      \caption{(color online) Distributions of the variables
	(remaining five of fifteen) used as inputs to the RF
	classifier for electron and muon channels combined, and before
	$b$-tagging.  The signal and background predictions and the
	systematic uncertainty band are evaluated after the fit of the
	total $WV$ cross section in the RF output distribution.
	Definitions for each variable are provided in the text 
        (LP denotes light partons such as $u$, $d$, $s$ 
        or gluon, and HF denotes heavy-flavor such as $c\bar{c}$ 
        or $b\bar{b}$).}
	\label{fig:rf_inputs2}
    \end{centering}
  \end{figure*}

\section{Systematic Uncertainties}

Table~\ref{tab:systEMMU} gives the size of the systematic
uncertainties for Monte Carlo simulations and multijet estimates.  We
consider the effect of systematic uncertainties both on the
normalization and on the shape of differential distributions for
signal and backgrounds.  Although Table~\ref{tab:systEMMU} lists 
uncertainties for the diboson and $W$+jets simulation, these 
uncertainties are not used when measuring the diboson signal cross 
section, for which the diboson and $W$+jets normalizations are  
free parameters.  However, the size of the uncertainty must be 
specified when estimating the significance and when we constrain 
the cross section for $WW$ production to its SM prediction in the fit.

  \begin{table*}[htb]

    \caption{The RMS amplitude (in percent) of each systematic
    uncertainty in the RF output distributions for the signal and
    background predictions.  The RMS amplitude is defined as:
    $\sqrt{\sum_{i=0}^{n}{p_i \Delta_i^2}/\sum_{i=0}^{n}{p_i}}$; where
    $p_i$ is the predicted number of events in bin $i$, $\Delta_i$ is
    the percent change in bin $i$ when the uncertainty is varied by 1~s.d., 
    and $n$ is the number of bins.  In cases where the amplitude
    is different for different subchannels, the range of amplitudes is
    given.  The rightmost column indicates whether the uncertainty
    only affects the normalization (N) or if it has also a differential
    dependence (D).}

    \label{tab:systEMMU}

    \begin{ruledtabular}
      \begin{tabular}{l @{\extracolsep{\fill}} 
      r @{\ \ \extracolsep{\fill}} l @{\extracolsep{\fill}\ \ } 
      r @{\ \ \extracolsep{\fill}} l @{\extracolsep{\fill}\ \ } 
      r @{\ \ \extracolsep{\fill}} l @{\extracolsep{\fill}\ \ } 
      r @{\ \ \extracolsep{\fill}} l @{\extracolsep{\fill}\ \ } 
      r @{\ \ \extracolsep{\fill}} l @{\extracolsep{\fill}\ \ } 
      c
      c}
	\multicolumn{1}{c}{Source of systematic}
	& \multicolumn{2}{c}{\multirow{2}{*}{Diboson signal}}
	& \multicolumn{2}{c}{\multirow{2}{*}{$W$+jets}}
	& \multicolumn{2}{c}{\multirow{2}{*}{$Z$+jets}}
	& \multicolumn{2}{c}{\multirow{2}{*}{Top}}
	& \multicolumn{2}{c}{\multirow{2}{*}{Multijet}}
	& \multicolumn{1}{c}{\multirow{2}{*}{Nature}}
	& \\
	\multicolumn{1}{c}{uncertainty}
	&&
	&&
	&&
	&&
	&&
	&
	&\\

	\hline
        Electron trigger/ID efficiency                        && $\pm $5    && $\pm $5    && $\pm $5    && $\pm $5    &&         & N& \\
        Muon trigger/ID efficiency                            && $\pm $5    && $\pm $5    && $\pm $5    && $\pm $5    &&         & N& \\
	Jet identification                                    && $\pm $1    && $\pm $1--2 && $\pm $1--2 && $\pm<$1--2 &&         & D& \\
	Jet energy scale                                      && $\pm $2--4 && $\pm $6--8 && $\pm $4--12&& $\pm $2--3 &&         & D& \\
	Jet energy resolution                                 && $\pm $2--3 && $\pm $3--12&& $\pm $4--10&& $\pm $1--2 &&         & D& \\
	Jet vertex confirmation                               && $\pm $2--3 && $\pm $3--4 && $\pm $3--5 && $\pm $1--3 &&         & D& \\
	Taggability correction                                && $\pm<$1    && $\pm<$1    && $\pm<$1    && $\pm<$1    &&         & D& \\
	$b$-tagging                                             && $\pm $1--5 && $\pm $1--4 && $\pm $1--5 && $\pm $8--10&&         & D& \\
	Luminosity                                            && $\pm $6.1  && $\pm $6.1  && $\pm $6.1  && $\pm $6.1  &&         & N& \\
	Cross section                                         && $\pm $7    && $\pm $6.3  && $\pm $6.3  && $\pm $10   &&            & N& \\
	$V+$heavy-flavor cross section                        &&            && $\pm $20   && $\pm $20   &&            &&            & N& \\
	$V$+2 jets/$V$+3 jets cross section                   &&            && $\pm $10   && $\pm $10   &&            &&            & N& \\
	Multijet normalization                                &&            &&            &&            &&            && $\pm $20   & N& \\
	Multijet shape, electron channel                      &&            &&            &&            &&            && $\pm<$1    & D& \\
	Multijet shape, muon channel                          &&            &&            &&            &&            && $\pm<$1    & D& \\
	Diboson modeling                                      && $\pm $2--3 &&            &&            &&            &&            & D& \\
        Parton distribution function                          && $\pm $1    && $\pm $2    && $\pm $1--3 && $\pm $2--4 &&            & D& \\ 
	Unclustered Energy correction                         && $\pm<$1    && $\pm<$1    && $\pm<$1    && $\pm<$1    &&            & D& \\
	{\sc alpgen} jet $\eta$ corrections                   &&            && $\pm<$1    && $\pm<$1    &&            &&            & D& \\
	{\sc alpgen} $\Delta R(jj)$ and $p_T(W)$ corrections  &&            && $\pm<$1    && $\pm<$1    &&            &&            & D& \\
	Re-weighting diboson bias                             &&            && $\pm<$1    && $\pm<$1    && $\pm<$1    &&            & D& \\
	Renormalization and factorization scales              &&            && $\pm<$1    && $\pm<$1    &&            &&            & D& \\ 
	Underlying event model                                &&            && $\pm<$1    && $\pm $1    &&            &&            & D& \\ 
	{\sc alpgen} parton-jet matching parameters           &&            && $\pm<$1    && $\pm<$1    &&            &&            & D& \\ 
      \end{tabular}
    \end{ruledtabular}
  \end{table*}

\section{$\Delta\chi^2$ for $WV$ Measurement}

The statistical significance of the diboson signal yield from the 
fit to the data is estimated via analysis of the $\Delta\chi^2$ 
curve obtained by fitting the data to the sum of background and 
signal templates as a function of the signal rate.  The results 
of this analysis are given in Table~\ref{tab:WV}.  Figure~\ref{fig:WVchi2} 
shows how the $\chi^2$ of the fit changes as a function of the signal cross 
section when using either the dijet mass or the RF output distribution 
to measure the total $WV$ cross section. 

  \begin{table*}[htb]

    \caption{ Results from fitting the total $WV$ cross section and
    the uncertainties resulting from limited data statistics (stat),
    and possible systematic biases (syst).  Also, the expected and 
    observed $\Delta\chi^2$ obtained by fitting the data with and 
    without the specified signal process and the corresponding 
    significance in number of standard deviations (s.d.)  for a 
    one-sided Gaussian integral.  }
   \label{tab:WV}
    \begin{ruledtabular}
      \begin{tabular}{lccc}
      & & \multicolumn{2}{c}{$\Delta\chi^2$ (significance)} \\
         & Measured $\sigma(WV)$ [pb] & Expected & Observed \\
 	\hline
	\vspace{1mm}
	 RF Output & 19.6~$\pm$~1.4~(stat)$~^{+2.9}_{-2.7}$~(syst) & \wvRFdchiExp~(\wvRFsdexp~s.d.) & \wvRFdchi~(\wvRFsd~s.d.) \\
	 Dijet Mass & 18.3~$\pm$~1.5~(stat)$~^{+3.5}_{-3.3}$~(syst) & \wvMJJdchiExp~(\wvMJJsdexp~s.d.) & \wvMJJdchi~(\wvMJJsd~s.d.)  \\
      \end{tabular}
    \end{ruledtabular}
  \end{table*}

  \begin{figure*}[hbt]
    \begin{centering}
      \includegraphics[width=4in]{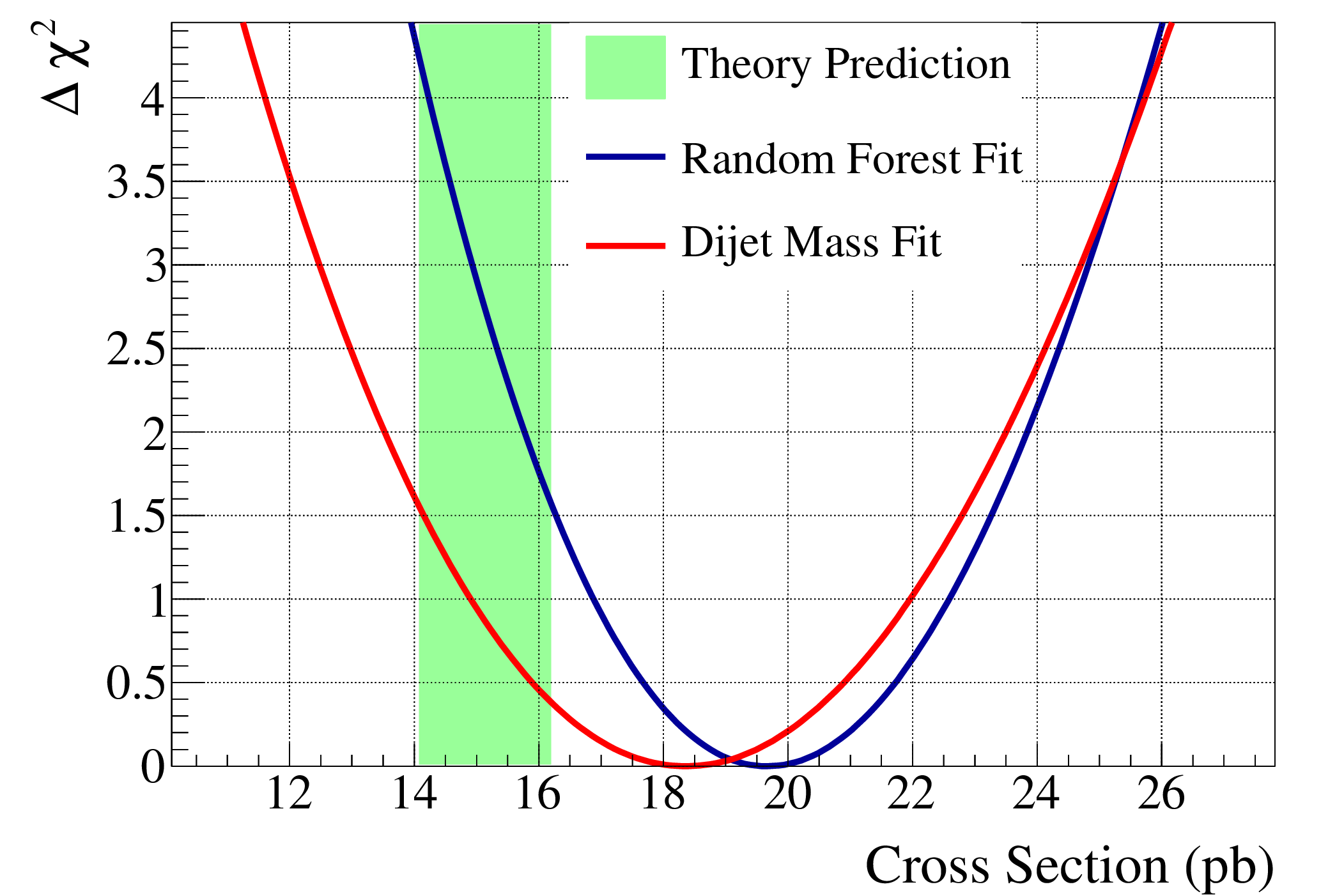}
    \caption{(color online) The change in $\chi^2$ relative to the best fit value
    when fitting the total $WV$ cross section using either the dijet
    mass or the RF output distribution. } \label{fig:WVchi2}
    \end{centering}
  \end{figure*}
	
\clearpage


\begin{thebibliography}{99}
\bibitem{bib:def}	     D0 uses a spherical coordinate system with the $z$ axis running along the proton beam axis. The angles $\theta$ and $\phi$ are the polar and azimuthal angles, respectively. Pseudorapidity is defined as $\eta = -\ln \left[\tan(\theta/2) \right]$, in which $\theta$ is measured with respect to the proton beam direction.
\end{thebibliography}
\end{document}